\documentclass{emulateapj}
\usepackage{epstopdf}
\usepackage{graphicx}
\tighten

\def\apjref#1;#2;#3;#4 {\par\pni\ #1,  #2, {\bf #3}, #4. \par}

\def\etal{{\it et al.~}}

\def\simlt{\lower.5ex\hbox{$\; \buildrel < \over \sim \;$}}
\def\simgt{\lower.5ex\hbox{$\; \buildrel > \over \sim \;$}}
\def\solar{\ifmmode _{\mathord\ odot}\else $_{\mathord\odot}$\fi}
\def\msun{\ifmmode {\rm M}_{\mathord\odot}\else $M_{\mathord\odot}$\fi}
\def\lsun{\ifmmode {\rm L}_{\mathord\odot}\else $L_{\mathord\odot}$\fi}
\def\xpvec{x\lla*p{\lower1ex\hbox{$\scriptstyle\sim$}}_{\scriptscriptstyle\perp}}
\def\xvec{x\llap{\lower1ex\hbox{$\scriptstyle\sim$}}}
\def\qvec{q\llap{\lower1ex\hbox{$\scriptstyle\sim$}}}
\def\to{\ifmmode \rightarrow\else $\rightarrow$\fi}
\def\Mc{M\raise0.5ex\hbox{c}}
\def\del{\nabla}

\def\none{\ifmmode ^{-1}\else $^{-1}$\fi}
\def\two{\ifmmode ^{2}\else $^{2}$\fi}
\def\ntwo{\ifmmode ^{-2}\else $^{-2}$\fi}
\def\three{\ifmmode ^{3}\else $^{3}$\fi}
\def\nthree{\ifmmode ^{-3}\else $^{-3}$\fi}
\def\four{\ifmmode ^{4}\else $^{4}$\fi}
\def\nfour{\ifmmode ^{-4}\else $^{-4}$\fi}
\def\five{\ifmmode ^{5}\else $^{5}$\fi}
\def\nfive{\ifmmode ^{-5}\else $^{-5}$\fi}

\def\g{\ifmmode {\rm g}\else g\fi}
\def\kg{\ifmmode {\rm kg}\else kg\fi}

\def\cm{\ifmmode {\rm cm}\else cm\fi}
\def\m{\ifmmode {\rm m}\else m\fi}
\def\km{\ifmmode {\rm km}\else km\fi}
\def\pc{\ifmmode {\rm pc}\else pc\fi}
\def\ly{\ifmmode {\rm ly}\else ly\fi}
\def\au{\ifmmode {\rm au}\else au\fi}

\def\s{\ifmmode {\rm s}\else s\fi}
\def\Hz{\ifmmode {\rm Hz}\else Hz\fi}
\def\y{\ifmmode {\rm y}\else y\fi}

\def\K{\ifmmode {\rm K}\else K\fi}
\def\ster{\ifmmode {\rm ster}\else ster\fi}
\def\erg{\ifmmode {\rm erg}\else erg\fi}
\def\dyn{\ifmmode {\rm dyn}\else dyn\fi}

\newcommand{\nbh}	{\bar n_{\rm H}}

\newcommand{\e}	{\ifmmode ^{-1}\else $^{-1}$\fi}
\newcommand{\ee}	{\ifmmode ^{-1}\else $^{-2}$\fi}
\newcommand{\eee}	{\ifmmode ^{-1}\else $^{-3}$\fi}
\def\calm	{{\cal M}}

\renewcommand{\baselinestretch}{1.0}\small\normalsize

\begin{document}

\title{The Kinematics of Molecular Cloud Cores in the Presence of Driven and Decaying Turbulence:
Comparisons with Observations}

\author{Stella S. R. Offner} \affil{Department of Physics, University of
California, Berkeley, CA 94720} 
\email{soffner@berkeley.edu}

\author{Mark R. Krumholz \altaffilmark{1}} \affil{Department of Astrophysical Sciences, Princeton University, Princeton, NJ 08544} 
\altaffiltext{1}{Hubble Fellow}
\email{krumholz@astro.princeton.edu}

\author{ Richard I. Klein}
\affil{Department of Astronomy, University of California, Berkeley 
CA 94720, USA, and Lawrence Livermore National Laboratory, P.0. Box 808, L-23,
Livermore, CA 94550}
\email{klein@astron.berkeley.edu}

\author{Christopher F. McKee}
\affil{Departments of Physics and Astronomy, University of California, Berkeley, CA 94720}
\email{cmckee@astron.berkeley.edu}

\begin{abstract}
In this study we investigate the formation and properties of prestellar and protostellar cores using hydrodynamic, self-gravitating Adaptive Mesh Refinement simulations, comparing the cases where turbulence is continually driven and where it is allowed to decay. We model observations of these cores in the C$^{18}$O$(2\rightarrow 1)$, NH$_3(1,1)$, and N$_2$H$^+(1\rightarrow 0)$ lines, and from the simulated observations we measure the linewidths of individual cores, the linewidths of the surrounding gas, and the motions of the cores relative to one another. Some of these distributions are significantly different in the driven and decaying runs, making them potential diagnostics for determining whether the turbulence in observed star-forming clouds is driven or decaying. Comparing our simulations with observed cores in the Perseus and $\rho$ Ophiuchus clouds shows reasonably good agreement between the observed and simulated core-to-core velocity dispersions for both the driven and decaying cases. However, we find that the linewidths through protostellar cores in both simulations are too large compared to the observations. The disagreement is noticably worse for the decaying simulation, in which cores show highly supersonic infall signatures in their centers that decrease toward their edges, a pattern not seen in the observed regions. 
This result gives some support to the 
use of driven turbulence for modeling regions of star formation, but
reaching a firm conclusion on the relative merits of driven or decaying turbulence will require more complete data on a larger sample of clouds as well as simulations that include magnetic fields, outflows, and thermal feedback from the protostars.

\end{abstract}

\section {Introduction}

The origin of the stellar initial mass function (IMF) is one of the most important problems in astrophysics. Since the discovery of supersonic linewidths in star forming regions, understanding turbulence has been crucial for developing the theoretical framework for molecular cloud (MC) evolution,  core formation, and the IMF. Ongoing debate in this field concerns whether the formation and destruction of MCs is dynamic and non-equilibrium (e.g. Elmegreen 2000; Hartmann 2001; Dib et al. 2007) or slow and quasi-equilibrium (Shu et al. 1987; McKee 1999; Krumholz et al. 2006b; Krumholz \& Tan 2007; Nakamura \& Li 2007). The former mode would be characterized by transient turbulence, dissipating quickly on timescales comparable to the cloud lifetime so that GMCs are destroyed within $\sim$ one dynamical time. The latter case corresponds to regenerated turbulence, perhaps injected by the formation of the cloud, protostellar outflows, H II regions, external cloud shearing or supernova blastwaves, that is sufficiently strong to inhibit global gravitational collapse over many dynamical times. As shown by Offner et al. (2008) and Krumholz et al. (2005), the presence or absence of turbulent feedback directly relates to the physical mechanism of star formation and determines whether stars form by the formation and collapse of discrete protostellar cores (Padoan \& Nordlund 2002; McKee \& Tan 2002) or competitive accretion (Bonnell at al. 2001). In the turbulent core model, the cloud remains near virial equilibrium on large scales and collapse occurs only locally in cores that are created and then mass-limited by the initial turbulent compressions. In the competitive accretion model, turbulence generates the initial overdensities, but without turbulent support, the cores are wandering accreting seeds, competing for gas from a reservoir, limited only by the size of the MC as a whole.

There have been a number of recent observational papers investigating starless and protostellar core velocity dispersions, envelopes, and relative motions (Andr\'{e} et al. 2007; Kirk et al. 2007; Muench et al. 2007; Rosolowsky et al. 2007; Walsh et al. 2004),  quantities that provide important clues about the core lifetimes and evolution, and about the turbulent state of the natal MC. All of these results, which include observations of a range of star forming regions in different tracers, indicate that observed low-mass cores have approximately sonic central velocity dispersions, at most transonic velocity dispersions in their surrounding envelopes, and relative motions that are slower than the virial velocity of the parent environment.  Such results potentially contradict core properties measured in simulations in collapsing clusters exhibiting competitive accretion (Ayliffe et al. 2007; Klessen et al. 2005; Tilley \& Pudritz 2004).

In this paper we analyze the simulations described in Offner et al. (2008), which follow the evolution of an isothermal turbulent molecular cloud with and without continuous injection of energy to drive turbulent motions. These simulations use the adaptive mesh refinement (AMR) code Orion (Truelove et al. 1998, Klein 1999). The goal of our present work is to explore differences between cores forming in these two environments and to provide predictions of their properties for observational comparison. For this purpose, we simulate observations of our cores using dust continuum and molecular lines, with realistic telescope resolutions.  Unlike earlier comparisons of isothermal self-gravitating simulations with observations (Ayliffe et al. 
2007; Klessen et al. 2005; Ballesteros-Paredes et al. 2003), we perform more detailed radiative post-processing in order to simulate more accurately synthetic observations of our data. We also compare these observational measures for both driven and decaying turbulence, which has not previously been investigated. 
Keto \& Field (2005) obtain post-processed simulated line profiles of several common tracers modeled with a non-LTE radiative transfer code and find good agreement with observed isolated cores. However, their initial conditions are simple 1-D non-turbulent hydrostatic models and they halt the calculations when the central cores velocity exceeds the sound speed. 
Further, we report core-to-core centroid velocity dispersions of the simulated cores, which has not previously been studied in turbulent simulations. Work by Padoan et al. (2001) comparing observed large scale gas motions with 128$^3$ fixed-grid isothermal, non-self-gravitating, MHD simulations found good agreement with the gas centroid velocity dispersion-column density relation. In our higher resolution simulations, we instead focus on the smaller physical scales of self-gravitating cores and their observed properties, and we neglect the effects of MHD.

In section 2, we describe our simulations in detail. Section 3 contains the methods of data analysis we use to simulate observations of our AMR data. In section 4, we present our results on the central core dispersions, relative motions, and dispersions of the surrounding core envelopes. In section 5 we present quantitative comparisons with observational data. Finally, section 6 contains our conclusions.

\section { Simulation Parameters}

As described in Offner et al. (2008), our two simulations are periodic boxes
containing an isothermal, non-magnetized gas that is initially not
self-gravitating. We first drive turbulent motions in the gas for two box crossing times, until the turbulence reaches statistical equilibrium, i.e. the power spectrum and probability density function shapes are constant in time. We adopt a 1-D Mach number of 4.9 (3-D Mach number of 8.5). At the time gravity is turned on, which we label $t$=0, our two simulations are identical.  In one simulation energy injection is halted and the turbulence gradually decays, while in the other turbulent driving is maintained so that the cloud remains in approximate virial equilibrium. The initial virial parameter is defined by
\begin{equation}
{{{5\sigma_{\rm 1D}^2 R}} \over{ GM}} = \alpha_{\rm vir} \simeq 1.67,
\end{equation}
where $\sigma_{\rm 1D}$ is the velocity dispersion, $M$ is the cloud mass, and $R=L/2$ is the cloud radius. 
We use periodic boundary conditions and 4 levels of refinement, which corresponds to an effective 2048$^3$ base grid for an equal-resolution, fixed-grid calculation.

Isothermal self-gravitating gas is scale free, so we give the key cloud properties as a function
 of fiducial values for the number density of hydrogen nuclei, 
$\bar n_{\rm H}$, and gas temperature, $T$. It is then easy to scale the simulation results to the 
astrophysical region of interest. 
For the adopted values of the virial parameter and Mach number,
the box length, mass, and 1-D velocity dispersion are given by
\begin{eqnarray}
L &=&  \mbox{2.9 } {T_1 }^{1/2} {\bar n_{\rm H, 3}}^{-1/2}  \mbox{ pc}\, ,
\label{eq:l} \\
M &=&  \mbox{865 } {T_1}^{3/2} {\bar n_{\rm H, 3}}^{-1/2} \mbox{ M}_{\odot} \, ,
\label{eq:m}\\
\sigma_{\rm 1D} &=&  \mbox{0.9 } {T_{1}}^{1/2} \mbox{ km s}^{-1}\, ,\\
t_{\rm ff} &=&  \mbox{1.37 } {\bar n_{\rm H,3}}^{-1/2} \mbox{ Myr}\, ,
\end{eqnarray}
where we have also listed the free-fall time for the gas in the box for completeness.

These equations are normalized to a temperature $T_1 = T / 10$~K and 
average hydrogen nuclei number density $\bar n_{\rm H,3} = 
\bar n_{\rm H} / (1 \times 10^3 $ cm$^{-3})$.
For the remainder of this paper, all results will
be given assuming the fiducial scaling values of 
$\bar n_{\rm H}= 1.1\times 10^{3}$ cm$^{-3}$  and $T=10$~K (Perseus) or $\bar n_{\rm H}= 2.0\times 10^{4}$ cm$^{-3}$  and $T=20$~K 
($\rho$ Ophiuchus; see \S 5) and assuming a mean particle mass of $\mu$=2.33$m_{\rm H}$. These conditions place $\rho$ Ophiuchus slightly 
above the observed linewidth-size relation (Solomon et al. 1987; Heyer \& Brunt 2004):
\begin{equation}
\sigma_{\rm 1D}=0.5 \left( {L \over {1.0 {\rm pc}}} \right)^{0.5} \mbox{km s}^{-1},
\label{eq:lws}
\end{equation}
where $L$ is the cloud length (we assume that Perseus lies on this relation--see
\S 5 below).

Note that this relation differs somewhat from the relation given by Heyer \& Brunt (2004) since the length scale determined from a Principal 
Component Analysis is smaller than the actual size of the region being sampled (see McKee \& Ostriker 2007).
These parameters may be adjusted to different conditions using equations (2)-(5). However, once we simulate an observation of the data for a given tracer, the scaling is fixed. Using these units, the minimum cell size is $\sim90$~AU 
and 280 AU for $\rho$ Ophiuchus and Perseus, respectively.

In the simulations, we introduce sink particles in collapsing regions that violate the Jeans condition (Truelove et al. 1997) at the finest AMR level (Krumholz et al. 2004), where we adopt a Jeans number of $J=0.25$. Cores that contain sink particles are analogous to observed protostellar cores, which contain a central source, while cores without sink particles can be considered prestellar. This distinction is an important one in some cases and we discuss some differences in the two simulations in \S 4. Note that due to our resolution and neglect of protostellar outflows, the sink particles represent a mass upper limit for any potentially forming protostar.

\section{Analysis}

Since our goal in this paper is to contrast the simulations and compare them with
observations, we must attempt to replicate an observer's view of our
simulation. Observations of core kinematics, such as those of
Andr\'{e} et al. (2007, henceforth A07), Kirk et al. (2007, henceforth K07), and Rosolowsky et al. (2007, henceforth R07), generally trace the gas mass
using dust continuum data and obtain velocity information by observing
the same region in one or more molecular tracers. We process our
simulations using a rough approximation of these techniques as
follows. First, we select a fiducial cloud distance of either 125 pc,
corresponding to the distance to the Ophiuchus star-forming MC, or 260 pc 
for comparisons with the Perseus MC.
Second, we select an appropriate telescope resolution of $26"$ or $31"$ FWHM,
corresponding to 0.02 pc and 0.04 pc at our adopted distances, and approximate the
telescope beam as Gaussian in shape. We perform all line fits assuming
 0.047 km s$^{-1}$ velocity resolution per channel. Increasing the velocity resolution further has little effect on the line fits. For simplicity we adopt the same
resolution for observations in dust continuum and in all molecular
tracers. Our fiducial resolution is typical of observations of core
kinematics (e.g. A07, K07, R07).

For the dust continuum observations, since our gas and dust are isothermal and
the simulation domain is everywhere optically thin at typical
observing wavelengths of $\sim 1$ mm, the dust intensity emerging from
a given pixel is simply proportional to the column density in that
pixel. We therefore define a dust continuum map by computing
the column density and convolving the resulting map with the beam. To
avoid introducing unnecessary and artificial complications, we neglect
observational uncertainties in the conversion from an observed
dust continuum intensity to a column density, and assume that
the column density can be reconstructed accurately except for beam
smearing effects. 
We identify cores by finding the local maxima directly from the column 
density data. In the analysis, we consider only local maxima with peak columns greater
than 0.1 of the global maximum column of the smeared data. This cutoff corresponds to 
$\sim$ twice the mean smeared column density.

To model molecular line observations, we choose three representative
lines, the $J=2\rightarrow 1$ transition of C$^{18}$O, $J=1\rightarrow 0$ 
transition of N$_2$H$^+$, and the
NH$_{3}(1,1)$ transition, which have critical densities of $4.7 \times 10^3$ cm$^{-3}$,  
$6.2 \times 10^4$ cm$^{-3}$, and $1.9 \times 10^3$ cm$^{-3}$, respectively. (For this
calculation and all the others presented in this paper, we use
molecular data taken from the Leiden Atomic and Molecular
Database\footnote{See http://www.strw.leidenuniv.nl/$\sim$moldata},
Schoier et al. 2005).
These lines are often used in observing core
kinematics because they span a range of densities and, with the
exception of C$^{18}$O along the densest sightlines, are generally
optically thin in low mass star-forming regions.
We discuss the issue of optical depths in more detail in \S~\ref{odepths}.

We generate a position-position-velocity (PPV) data cube from
our simulations in each of these lines using a two step procedure,
which combines elements of
Krumholz et al. (2007a) and Krumholz et al. (2007b).
The first step is to compute the emissivity
as a function of density. Since, as we shall see, the
density-dependence of the molecular emission has important
consequences, we cannot assume that these species are in local
thermodynamic equilibrium (LTE). Instead, we assume that the gas is in
statistical equilibrium, that it is optically thin, and that radiative
pumping by line photons is negligible. Note that the advection time 
of the gas is large compared to
the molecular collisional and radiative time scales, which are on
the order of a few years for the mean density of our
simulations. Thus, the gas reaches statistical equilibrium essentially
instantaneously relative to the gas motion.
Collisional excitation dominates over radiative excitation or
de-excitation by line photons along lines of sight where the
density is above the transition critical density. Since we are particularly
interested in the high density regions of the cores, we need not
consider radiative pumping in our analysis. However, we do include
radiative excitation and de-excitation due to the cosmic microwave
background, since this can be significant for lines at very low
frequencies such as NH$_3(1,1)$.

For a molecule like C$^{18}$O with no hyperfine structure, under these
approximations the fraction $f_i$ of molecules of a given species in
bound state $i$ is given by the equations of statistical equilibrium
\begin{eqnarray}
\label{stat_equilibrium}
\nonumber \sum_j (n_{\rm H_{\rm 2}} q_{ji} + A_{ji} + B_{ji} I_{\rm CMB}) f_j  \\
 = \left[\sum_k (n_{\rm H_{\rm 2}} q_{ik} + A_{ik} + B_{ik} I_{\rm CMB})\right] f_i \\
\label{stat_normalization}
\sum_i f_i & = & 1,
\end{eqnarray}
where $n_{\rm H_{\rm 2}}$ is the molecular hydrogen number density,
$q_{ij}$ is the collision rate for transitions from state $i$ to state
$j$, $A$ and $B$ are the Einstein coefficients for this
transition, and $I_{\rm CMB}$ is the intensity of the cosmic microwave
background radiation field (which is simply the Planck function for a
2.73 K blackbody) evaluated at the transition frequency. In this
expression we adopt the convention that the summations run over all
bound states, the spontaneous emission coefficient $A_{ij}=0$ for
$i\leq j$, that $B_{ij}$ is the stimulated emission coefficient for
$i>j$, the absorption coefficient for $i<j$, and is zero for $i=j$,
and that $q_{ij}=0$ for $i=j$. For molecules with hyperfine structure,
we show in Appendix \ref{hfappendix} that with some additional
approximations equation (\ref{stat_equilibrium})
continues to hold provided that the rate
coefficients $q_{ij}$, $A_{ij}$, and $B_{ij}$ are understood as
statistically-weighted sums over all the hyperfine sublevels of
states $i$ and $j$.

For molecules without hyperfine structure, 
the net emission minus absorption of the background CMB produced by a
parcel of gas along the line of sight is then given by
\begin{eqnarray}
\label{netintensity}
\nonumber j_{ij} - \chi_{ij} I_{\rm CMB} = \frac{h \nu_{ij}}{4\pi} X n_{\rm H}\\
\times [f_i (A_{ij} + B_{ij} I_{\rm CMB}) - f_j B_{ji} I_{\rm CMB}],
\end{eqnarray}
where $\chi_{ij}$ is the extinction of the CMB due to resonant
absorption, $\nu_{ij}$ is the transition frequency, $X$ is the
abundance
of the species in question relative to hydrogen nuclei, and $n_{\rm H}$ is
the number density of hydrogen nuclei. Physically, this quantity
represents the net radiation intensity added by transitioning
molecules over and above what one would see at that frequency due to
the CMB alone, under the assumption that the line is sufficiently thin
that the CMB dominates the intensity at that frequency. It is the
intensity one will observe in a line after subtracting off the
continuum. In the case of a molecule with hyperfine structure, under
the standard assumption that the hyperfine sublevels are populated in
proportion to their statistical weight (see Appendix
\ref{hfappendix}), the intensity produced by a
single transition from level $i$, hyperfine sublevel $\alpha$ to level
$j$, hyperfine sublevel $\beta$ is given by
\begin{eqnarray}
\label{hfnetintensity}
\nonumber j_{i\alpha j\beta} - \chi_{i\alpha j\beta} I_{\rm CMB} = 
\frac{h \nu_{i\alpha j\beta}}{4\pi} X n_{\rm H} \\
\left[f_i \frac{g_{i\alpha}}{g_i} (A_{i\alpha j\beta} + B_{i\alpha
j\beta} I_{\rm CMB}) - \frac{g_{j\beta}}{g_j} f_j B_{j\beta i\alpha}I_{\rm CMB} \right],
\end{eqnarray}
where $g_{i\alpha}$ is the statistical weight of hyperfine sublevel
$\alpha$, $g_i=\sum_{\alpha} g_i$ is the summed statistical weight of
all the hyperfine sublevels of state $i$, and the combination of
subscripts $i\alpha j\beta$ indicates the frequency or radiative
coefficient for transitions from level $i$, hyperfine sublevel
$\alpha$ to level $j$, hyperfine sublevel $\beta$. If one
neglects the very small differences in frequency between
the different hyperfine transitions (i.e.\ one takes $\nu_{i\alpha
j\beta}\approx \nu_{ij}$ independent of $\alpha$ and $\beta$) and
sums equation (\ref{hfnetintensity}) over hyperfine substates $\alpha$
and $\beta$, then it immediately reduces to equation
(\ref{netintensity}) provided that the rate coefficients are
understood to be statistically-weighted sums of the individual
hyperfine transition coefficients (per equations \ref{hfsums1} -
\ref{hfsums3}). Thus equation (\ref{netintensity}) gives the total
intensity summed over all hyperfine components. In either the presence
or absence of hyperfine splitting, to compute the intensity from our
simulations, we solve the system of equations
(\ref{stat_equilibrium})-(\ref{stat_normalization}) for our fiducial
temperature $T$ for a wide range of molecular densities $n_{\rm
H_2}$ and tabulate the quantities $(j_{ij} - \chi_{ij} I_{\rm CMB})/X$
or $(j_{i\alpha j\beta } - \chi_{i\alpha j\beta} I_{\rm CMB})/X$
as a function of $n_{\rm H_2}$.

The second step to generate the PPV cube from the simulation data is
to compute the emergent intensity in each pixel in each velocity
channel using our tabulated net emission function. The
specific emissivity minus absorption of the gas at a frequency $\nu$
is $(j_{ij} - \chi_{ij} I_{\rm CMB}) \phi(\nu)$ or $(j_{i\alpha
j\beta} - \chi_{i\alpha j\beta} I_{\rm CMB}) \phi(\nu)$, in the
absence or presence of hyperfine splitting,
where $\phi(\nu)$ is the line shape function. To determine
$\phi(\nu)$, we assume
that the molecules in each cell have a Maxwellian velocity
distribution with dispersion $\sigma_v = \sqrt{k_B T/m}$, where $m$
is the mass of the emitting molecule.
For this velocity distribution, the line shape function for a fluid
with bulk velocity $v$ is
\begin{equation}
\label{lineprof}
\phi(v_{\rm obs}; v) = \frac{1}{\sqrt{2\pi\sigma_{\nu}^2}}
\exp \left[ -\frac{(v-v_{\rm obs})^2}{2\sigma_{\nu}^2}
\right],
\end{equation}
where an observation at velocity $v_{\rm obs}$
is understood to mean an observation at frequency
$\nu = (1-v_{\rm obs}/c)
\nu_{ij}$ and where $\sigma_{\nu} = (\sigma_v/c) \nu_{ij}$. For optically
thin emission with no hyperfine structure at an observed velocity
$v_{\rm obs}$, a cell of length $\Delta x$ contributes a specific
intensity above the continuum of
\begin{equation}
I_{\nu} = (j_{ij}-\chi_{ij} I_{\rm CMB}) \Delta x \phi(v_{\rm obs}; v),
\end{equation}
where $j_{ij}$ and $\chi_{ij}$ are functions of the cell density
$n_{\rm H}$ and $\phi(v_{\rm obs}; v)$ is a function of the cell
velocity $v$. The intensity averaged over a velocity channel that
covers velocities in the range $v_0 \leq v_{\rm obs} \leq v_1$ is
\begin{eqnarray}
 \nonumber \left<I_{\nu}\right>_{\rm chan} = (j_{ij} - \chi_{ij} I_{\rm CMB})
\frac{c \Delta x}{4(v_1-v_0)\nu_{ij}} \\
\times \left[\mbox{erf}\left(\frac{v_1-v}{\sqrt{2} \sigma_v}\right)
- \mbox{erf}\left(\frac{v_0-v}{\sqrt{2} \sigma_v}\right)
\right].
\end{eqnarray}
We compute the channel-averaged specific intensity along each line of
sight by summing $\left<I_{\nu}\right>_{\rm chan}$
over all the cells, each with its own velocity $v$,  along the line of
sight. The final step in constructing our PPV data cube is that we take the
summed intensity computed in this way and smear each velocity channel
using our Gaussian beam.

In the case of molecules with hyperfine
structure, the equations are identical except that the subscripts $ij$
are replaced by $i\alpha j\beta$, and we note that, since the
hyperfine components are closely spaced in frequency, multiple
components may contribute significant intensity at the same
frequency. However, in the observations to which we wish to compare
our simulations, kinematic information is generally obtained by
fitting one or more well-separated individual hyperfine components
(e.g.\ A07, K07, although see R07, who use a more complex
procedure). Thus, in practice it is generally not necessary for our
purposes to consider more than a single hyperfine component. For
optically thin emission in hyperfine components with no significant
line overlap, this means that the procedures for molecules with and
without hyperfine splitting are the same.

Our procedure determines the emission only 
up to the unknown
abundance $X$, which in reality will depend on the emitting species
and on the density and
temperature, and probably also the thermal and density history, of a
given fluid element. For example, observations show that in the densest
cold regions CO and its isotopomers will be depleted, while the
abundance of N$_2$H$^+$ stays roughly constant (Tafalla et al.\ 2004a,b).
In order to approximate this effect, we adopt a simple depletion model 
for each of the chemical species that we
simulate. For C$^{18}$O, we assume an abundance of $X=10^{-7}$ molecules per H$_2$ molecule with depletion occurring at  $n_{\rm H_2}=5\times10^4$ cm$^{-3}$ (Tafalla et al. 2004a). For N$_2$H$^+$, we adopt $X=10^{-10}$ with depletion at  $n_{\rm H_2}=5\times10^7$ cm$^{-3}$ (K07; Tafalla et al. 2002). Although depletion in nitrogenous species is not generally observed, it is assumed that N$_2$ begins to disappear at number densities $n_{\rm H_2} > 10^6$ cm$^{-3}$ (Walmsley et al. 2003). For the NH$_3$ measurements we compare to in Perseus, we use $X= 10^{-8}$ (Rosolowsky, private communication) with assumed depletion at the same density as N$_2$H$^+$.

\begin{figure*}
\plotone{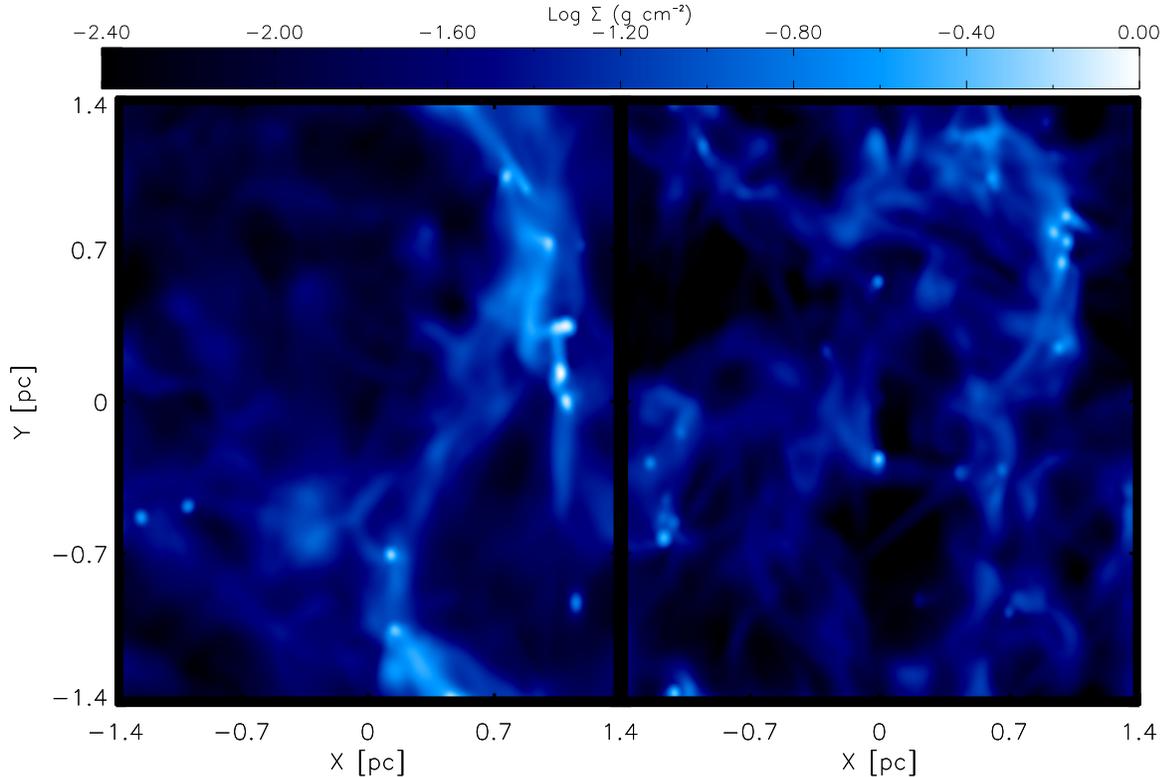} 
\figcaption{The images show the decaying (left) and driven (right) log column densities (g cm$^{-2}$) `observed' at a distance of 260 pc with beam size of 31''.
\label{col}}
\end{figure*}

We use these procedures to produce dust continuum / column density maps
and PPV cubes for each of our three molecular lines. To increase our
statistics, we generate data sets for each cardinal direction at
$t=t_{\rm ff}$, and we treat the three orientations as independent
observations. Figure \ref{col} shows a dust continuum map
in one particular orientation. 

\section{Results}

In the decaying simulation, at 1$t_{\rm ff}$ we identify a total of 109 cores, 54 of which can be considered protostellar due to the presence of a sink particle within 0.1 pc of the core center. In the driven simulation, we find 214 cores, 92 of which are protostellar. A large central point mass can have a significant effect on the core gas motion, so we separate out the `starless' cores for comparison. The relative number of starless cores to protostellar cores varies from star-forming cloud to cloud depending upon the advancement of star-formation in the region. The ratios of prestellar to protostellar cores that we find in our simulations are similar to the ratios observed in Perseus and Ophiuchus (Young et al. 2006; Enoch et al. 2006). In these simulations, the larger number of cores in the driven run is significant because the ongoing turbulence creates more new condensations, which also collapse more slowly.

For the sake of clarity, we will refer to the centroid velocities of the cores as the ``first moments" and the velocity dispersions through the core centers as the ``second moments." Thus in the following sections we will describe the measured distributions of the first and second moments and report the dispersion of the first moments (i.e. the core-to-core velocity dispersion). We define transonic velocities as those falling in the range $c_{\rm s} \leq \sigma  \leq 2c_{\rm s}$, while supersonic dispersions have $\sigma > 2c_{\rm s}$.

\subsection{Central Velocity Dispersions}

In this section, we investigate the distribution of second moments (central non-thermal velocity dispersions through the core centers) in N$_2$H$^+$, a measure that is useful for determining the level of turbulence and infall motion in the core. The total dispersion along the line of sight is given by
\begin{equation}
\sigma_{\rm LOS}=\sqrt{\sigma_{\rm NT}^2+\sigma_{\rm T}^2},
\end{equation}
where $\sigma_{\rm T}=\sqrt{k_{\rm B}T/m}$ and $\sigma_{NT}$ is the non-thermal component that we discuss here.  

\begin{deluxetable*}{cccccccc} 
\epsscale{0.8}
\tablewidth{0pt}
\tablecolumns{8}
\tablecaption{Central velocity dispersion median and mean for the two environments and core types at $1.0t_{\rm ff}$ in N$_2$H$^+$ normalized to the conditions in Perseus. \label{table1}}
\tablehead{
\colhead{} &
\multicolumn{3}{c}{Decaying} &
\colhead{} &
\multicolumn{3}{c}{Driven} \\
\colhead{} & \colhead{All} &
\colhead{Prestellar} & \colhead{Protostellar} &
\colhead{} &
\colhead{All} & \colhead{Prestellar} &
\colhead{Protostellar} 
}
\startdata
$N_{\rm cores}$ & 109 & 55 & 54 &   &214 & 122 & 92 \\ 
Median $\sigma_{\rm NT}/c_{\rm s}$ & 1.0 &	0.6	 &    2.9	&	& 1.1		&  0.9 & 2.1 \\ 
Mean $\sigma_{\rm NT}/c_{\rm s}$ &   2.2	& 	0.6	 &    3.8	&	& 1.8  	&1.2  & 2.7 
\enddata
\end{deluxetable*}

\begin{figure*}
\epsscale{0.8}
\plotone{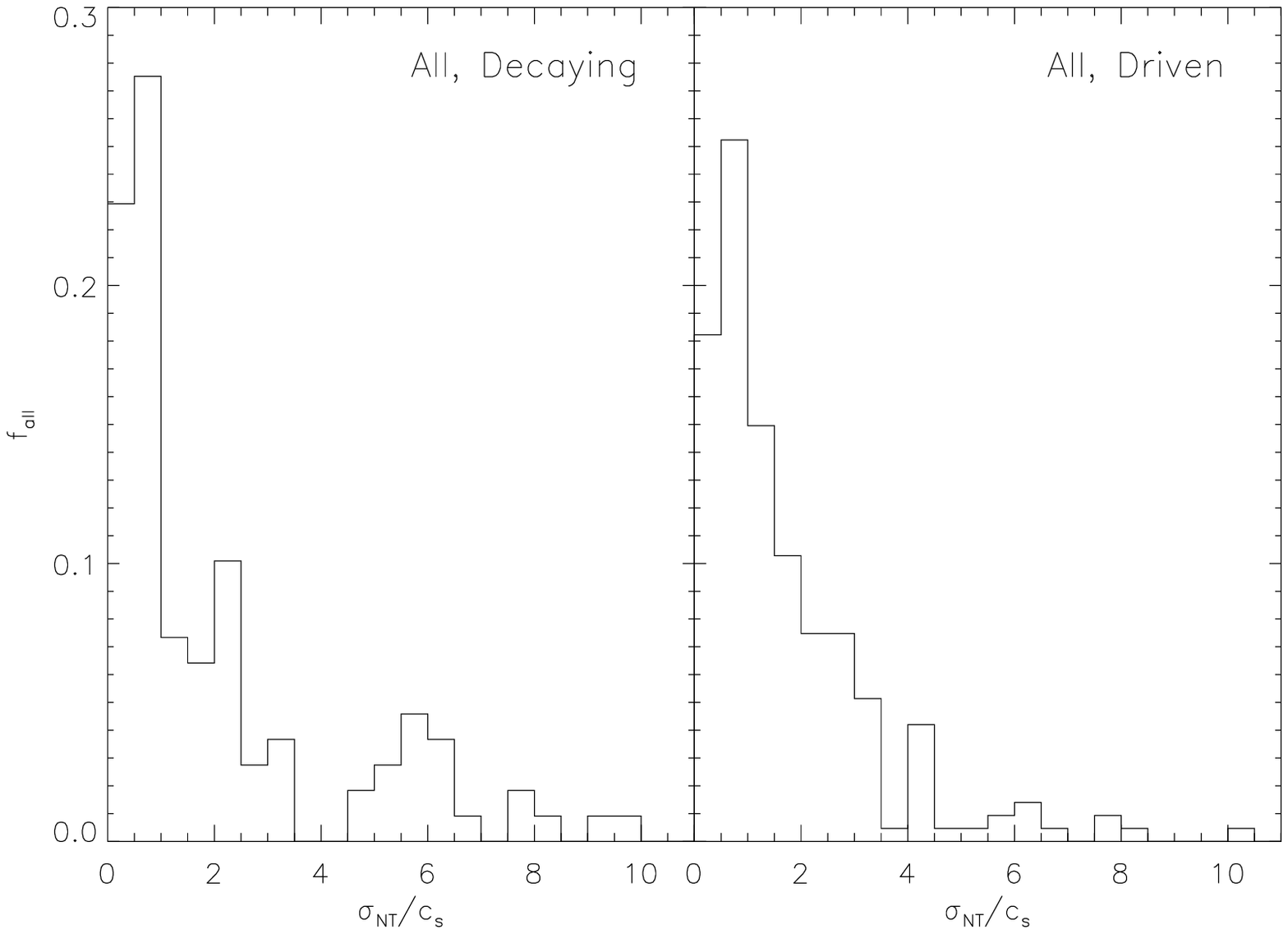}
\figcaption{Fraction $f$ of all cores binned as a function of second moments (non-thermal velocity
dispersion), $\sigma_{\rm NT}$, for a simulated observation of Perseus using N$_2$H$^+$.
The distribution on the left shows the cores in the decaying turbulence enviornment, while the distribution on the right gives the cores in the driven turbulence enviornment. \label{allvdisp}}
\end{figure*}

\begin{figure*}
\epsscale{0.8}
\plotone{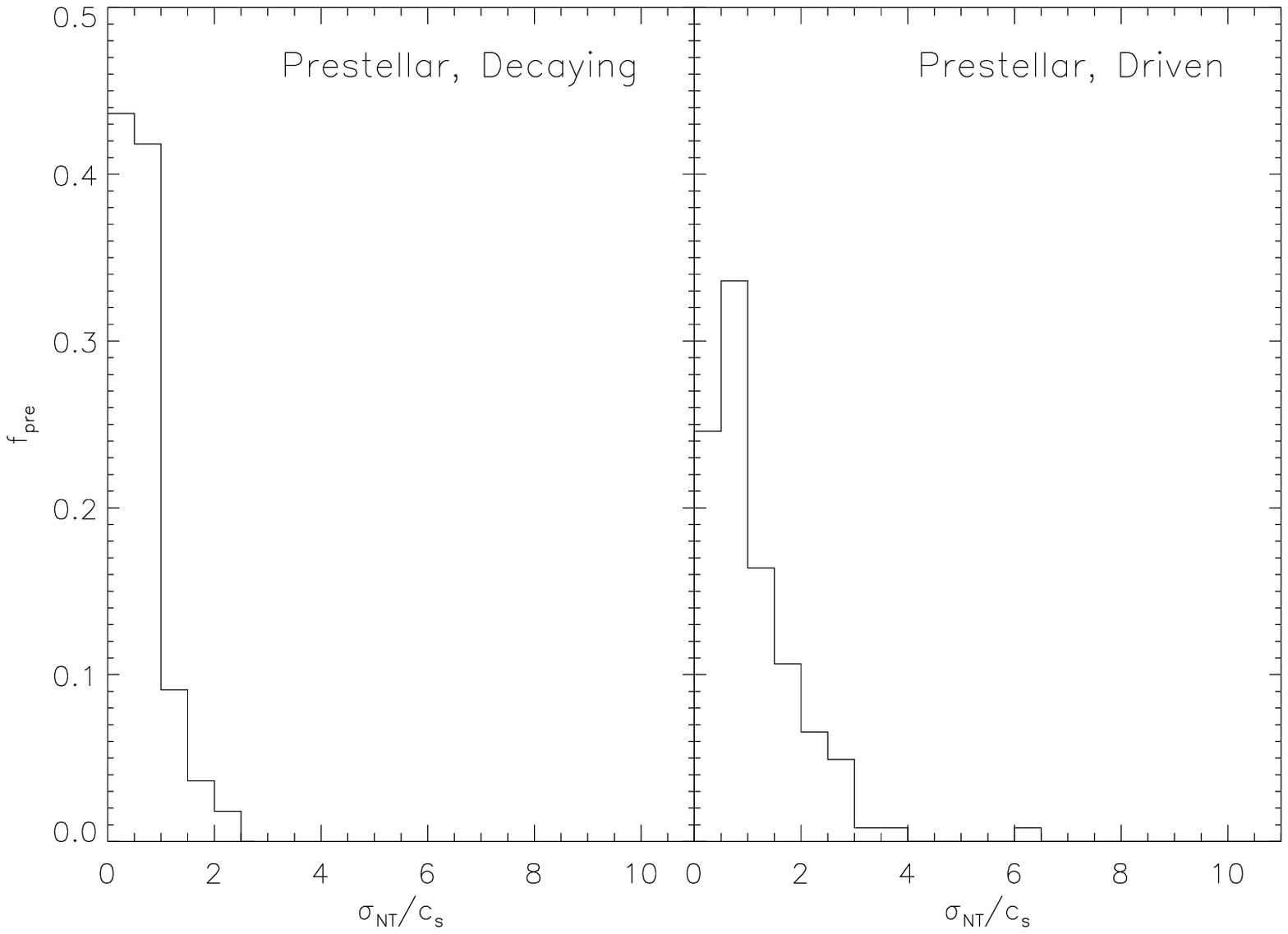}
\figcaption{Fraction $f$ of starless cores binned as a function of second moments (non-thermal velocity
dispersion), $\sigma_{\rm NT}$, for a simulated observation of Perseus using N$_2$H$^+$. The distribution on the left shows those cores in the decaying turbulence enviornment, while the distribution on the right gives the cores in the driven turbulence enviornment. \label{prevdisp}}
\end{figure*}

\begin{figure*}
\epsscale{0.8}
\plotone{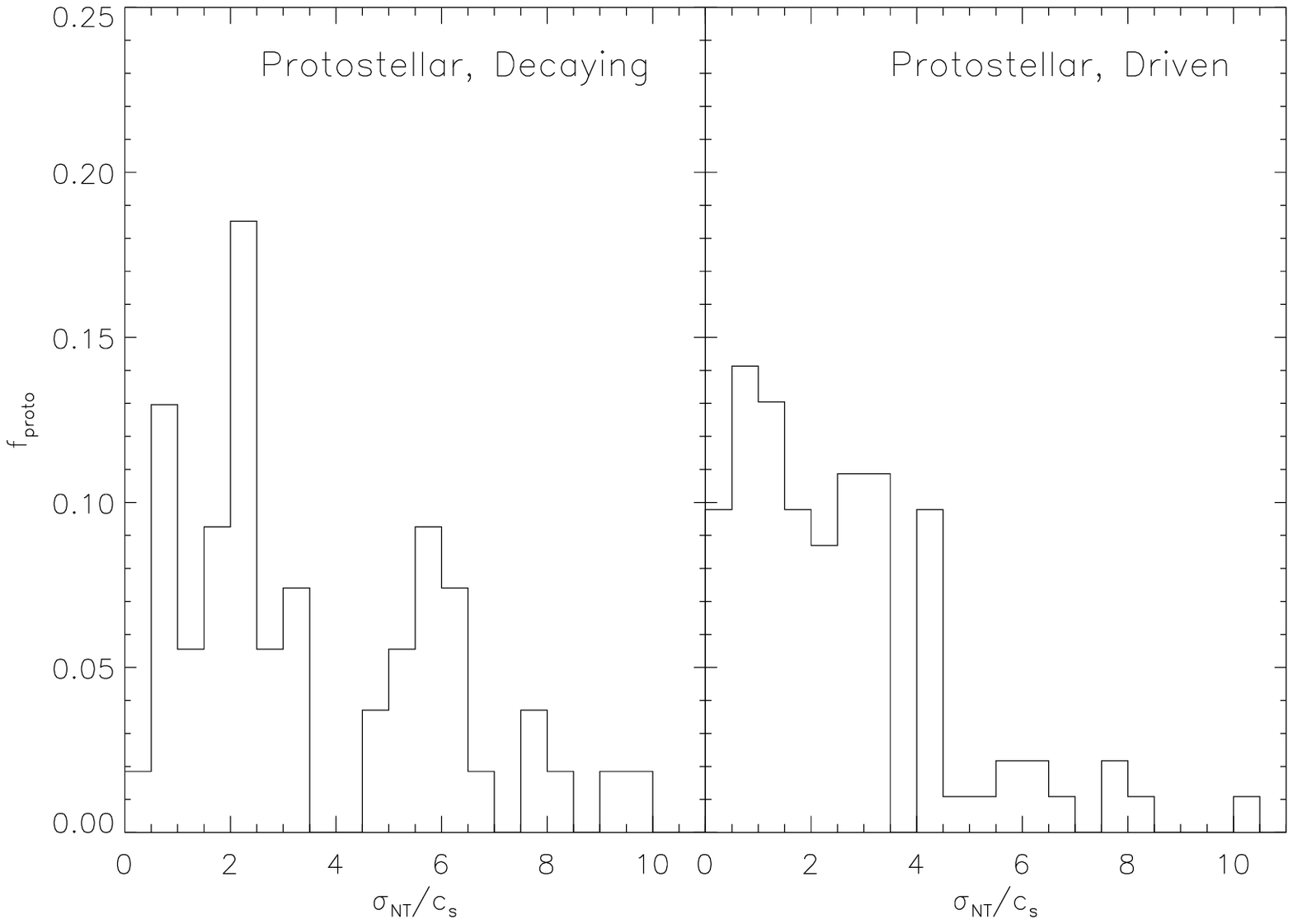}
\figcaption{Fraction $f$ of protostellar cores binned as a function of second moments (non-thermal velocity
dispersion), $\sigma_{\rm NT}$, for a simulated observation of Perseus using N$_2$H$^+$. The distribution on the left shows the cores in the decaying turbulence enviornment, while the distribution on the right gives the cores in the driven turbulence enviornment. \label{protovdisp}}
\end{figure*}

We compute $\sigma_{\rm LOS}$ in the simulations by fitting a Gaussian to the spectrum through the core center and then deriving the second moment, $\sigma_{\rm NT}$, from equation 14. Table \ref{table1} gives the median and means of $\sigma_{\rm NT}/c_{\rm s}$, and we plot the total distribution in figure \ref{allvdisp} and the prestellar and protostellar distributions in figures \ref{prevdisp}  and \ref{protovdisp}, respectively. The  core populations appear fairly similar in the two simulations, although there is evidence of the increased turbulence in the driven simulation. Since the cores are created by turbulent compressions in both environments, at early times they should have similar second moments. However, at late times, as the cores collapse and form protostars the distributions are more dissimilar. Indeed, from figure \ref{protovdisp} we can see that the protostellar distributions are much broader and less peaked than the prestellar ones. The decaying protostellar core population has almost twice as many cores in the tail ($\sigma_{\rm NT} > 4c_{\rm s}$) of the distribution, while the protostellar driven population is dominated by cores with $\sigma_{\rm NT} < 4c_{\rm s}$.

To better characterize the differences between the two simulations, we perform a Kolmogorov-Smirnov (KS) test comparing each of the core distributions. The KS statistic gives 1 minus the confidence level at which the null hypothesis that the two samples were drawn from the same underlying distribution can be ruled out, e.g. a KS statistic of 0.01 means that we can reject the hypothesis that the two samples were drawn from the same distribution at the 99\% confidence level. We find that the net driven and decaying velocity dispersion populations have a KS statistic of 18\%, meaning that we can rule out the hypothesis that they were drawn from the same population only with 82\% confidence. Individually, there is large disagreement in both the protostellar populations ($4\times 10^{-2}$\%) and prestellar core populations ($2\%$). 

The difference between the protostellar populations in the two simulations is associated with the mass differences between the sink particles: The decaying simulation has a median sink mass that is approximately twice that of the driven simulation and correspondingly larger accretion rates that are associated with higher velocity dispersions.

\subsection{Core Envelopes and Surroundings}

The velocity dispersions of the gas surrounding the central column density maxima yield information about the relative motion between core and envelope, and may also reveal the presence of shocks or strong infall that could limit core boundaries. Typically, observers find only small differences in velocity between the core and the surrounding gas envelope, which rules out dynamical pictures of core accretion in which protostars may strongly gravitationally interact with their neighbors (K07). In addition, although shocks are postulated to be the origin of the original density compression, close observations have not revealed strong confining shocks surrounding the cores.  Generally, our simulations produce prestellar cores that agree with the expectations from observations. However, the decaying protostellar cores exhibit supersonic internal velocities that are not observed in the star-forming regions we compare with. 

\begin{figure*}
\epsscale{0.7}
\plotone{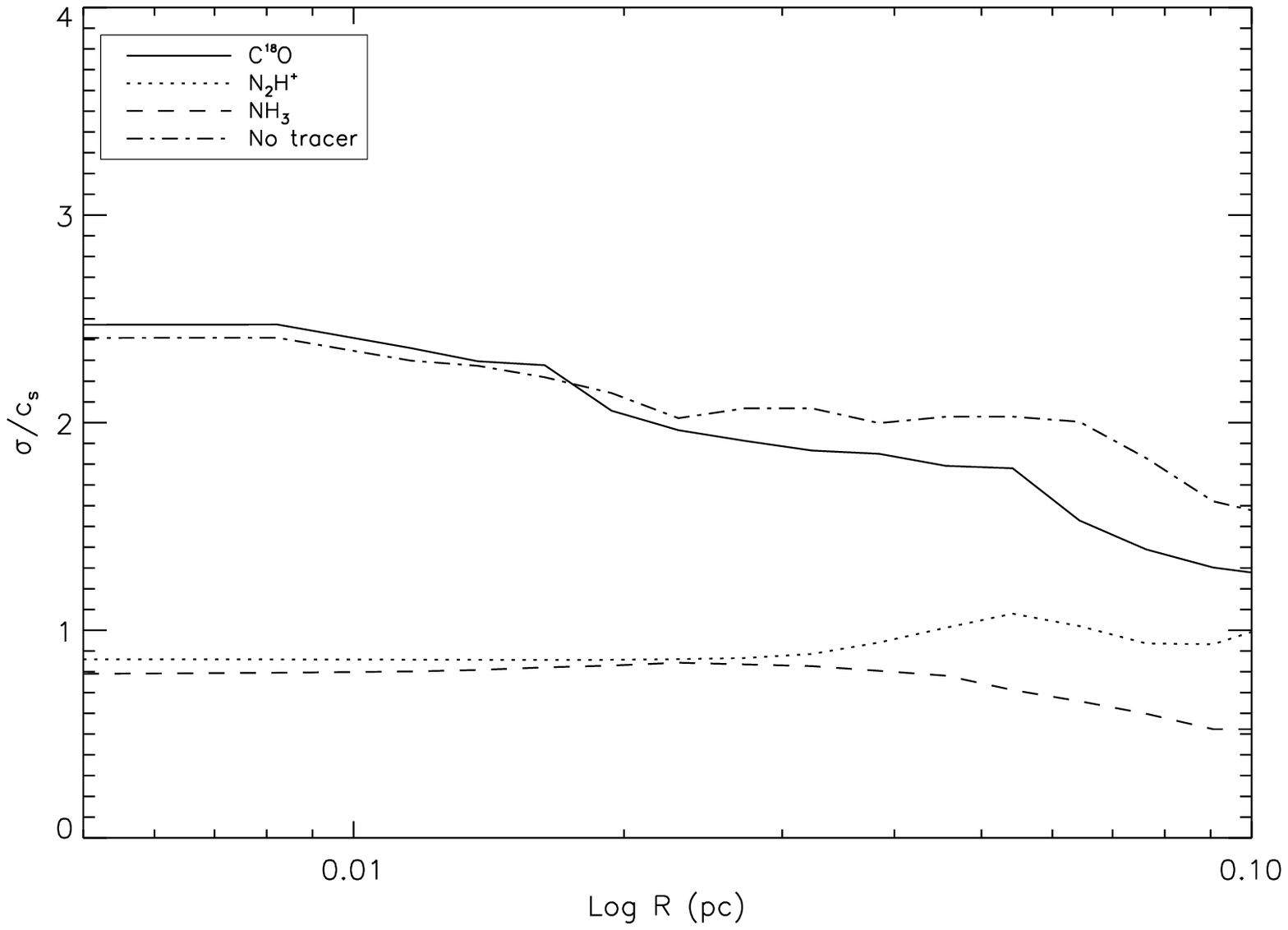}
\epsscale{0.8}
\plotone{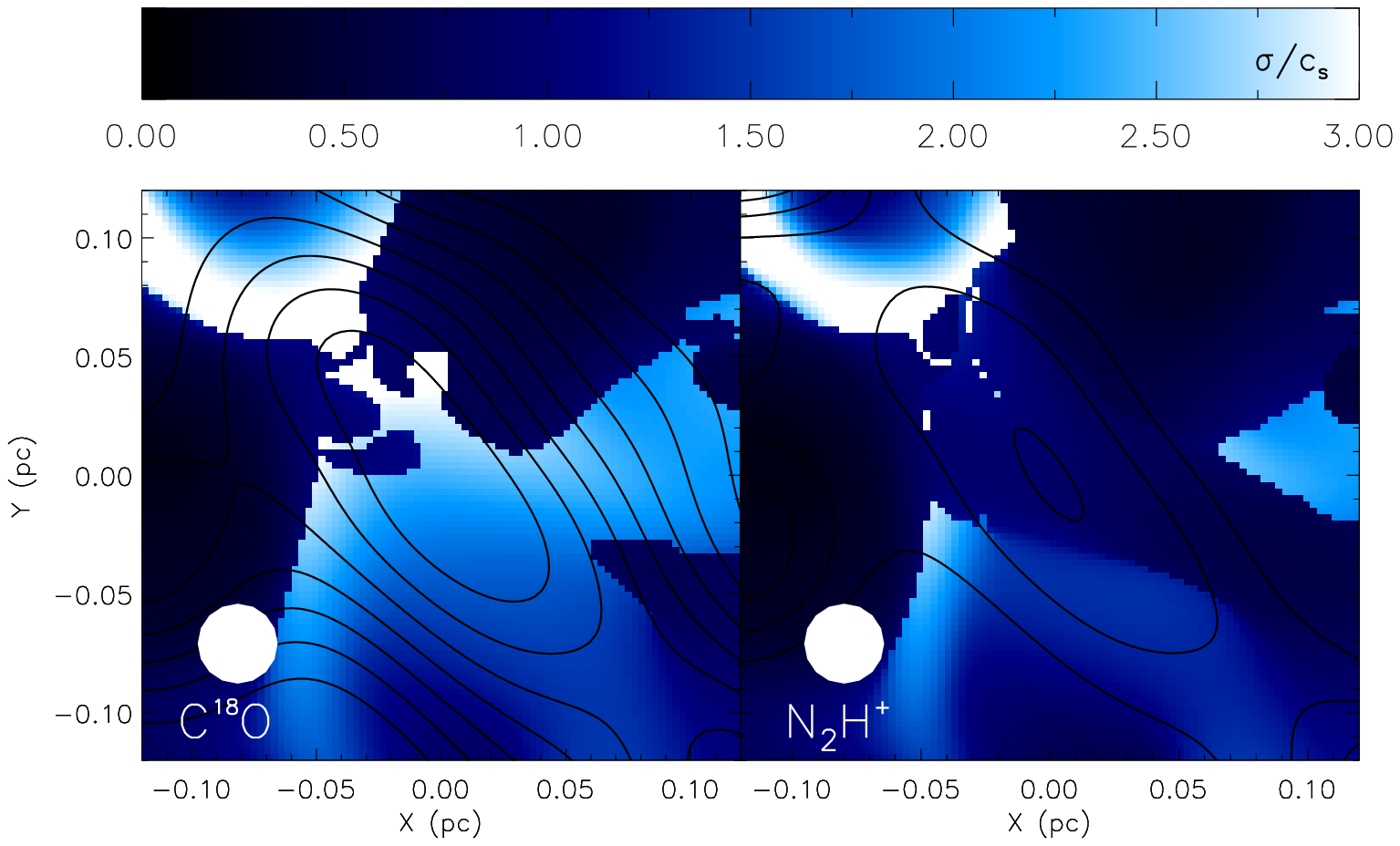}
\figcaption{ The upper plot gives average velocity dispersion as a function of radius for a single decaying starless core at $1t_{\rm ff}$. The images below show a simulated observation in C$^{18}$O (left) and N$_2$H$^+$ (right). Contours indicate integrated intensity where each contour is a 10\% linear change from the peak specific intensity in that tracer. The color scale shows velocity dispersion, $\sigma_{\rm NT}/c_{\rm s}$ and the circle indicates the FWHM beam size.
\label{dtCO_N2H+}}
\end{figure*}

\begin{figure*}
\epsscale{0.7}
\plotone{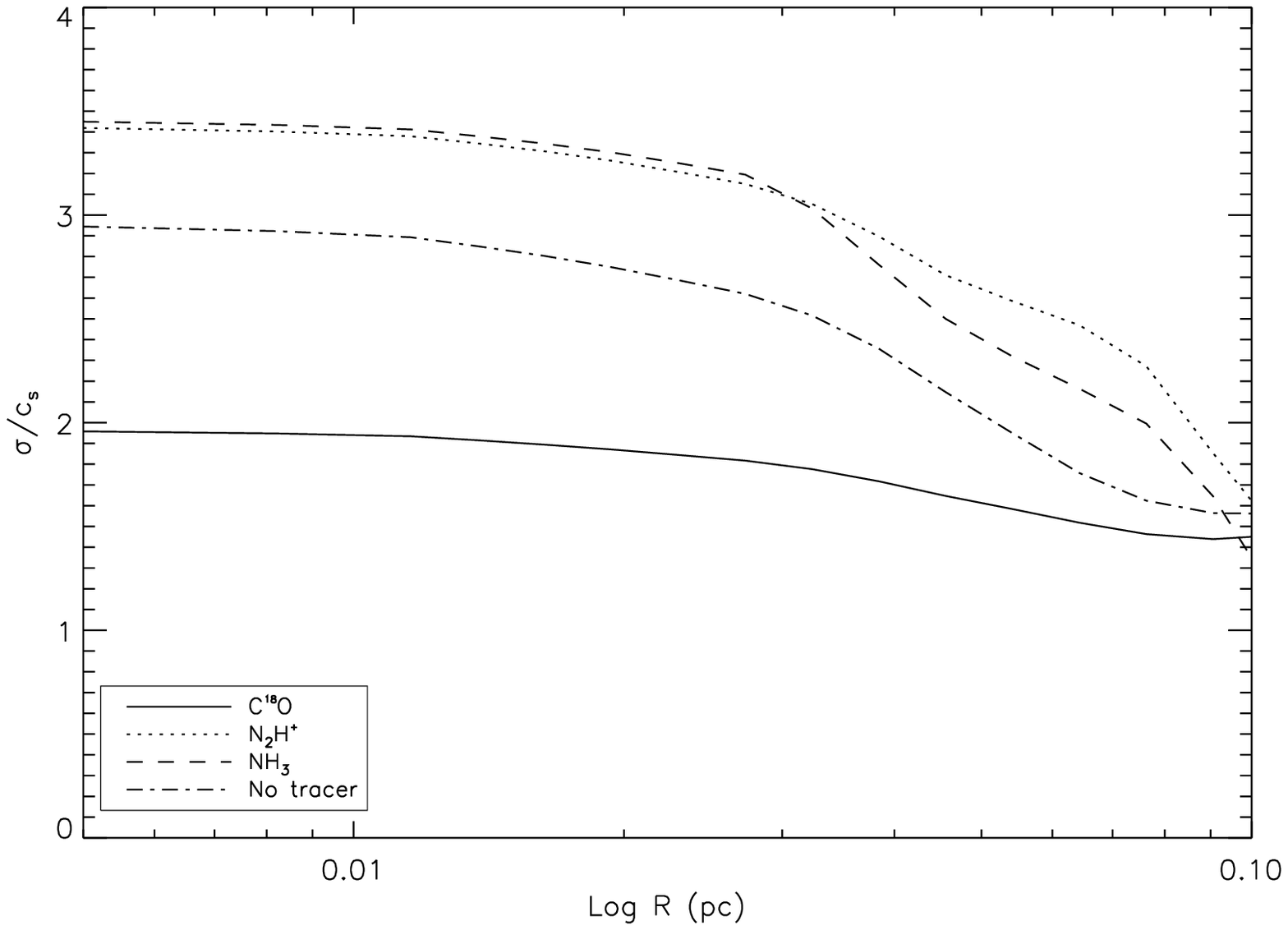}
\epsscale{0.8}
\plotone{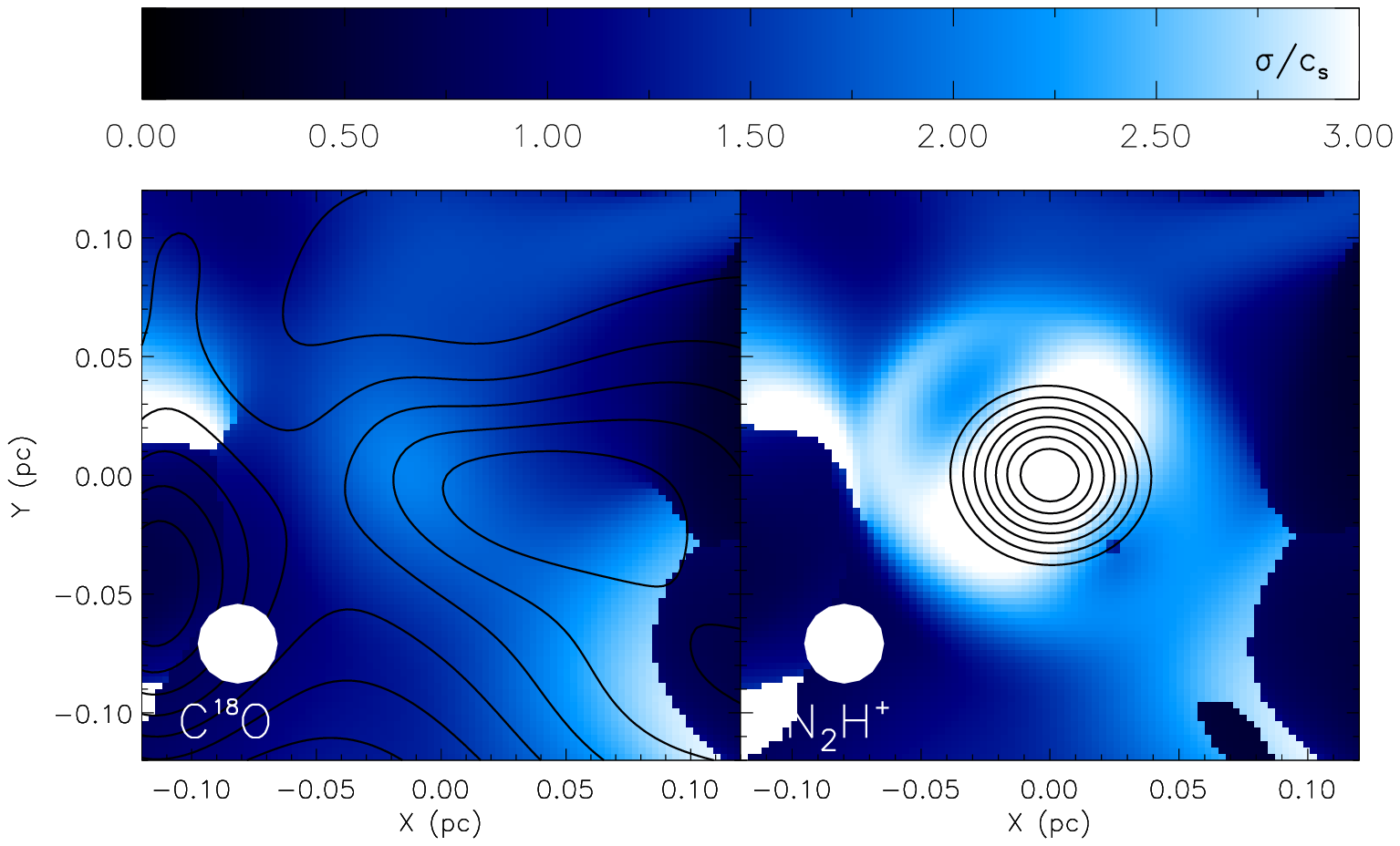}   
\figcaption{
The upper plot gives average velocity dispersion as a function of radius for a single decaying protostellar core at $1t_{\rm ff}$. The images below show a simulated observation in C$^{18}$O (left) and N$_2$H$^+$ (right). Contours indicate integrated intensity where each contour is a 10\% linear change from the peak specific intensity in that tracer. The color scale shows velocity dispersion, $\sigma_{\rm NT}/c_{\rm s}$  and the circle indicates the FWHM beam size. 
\label{vpCO_N2H+}}
\end{figure*}

\begin{figure*}
\epsscale{0.8}
\plotone{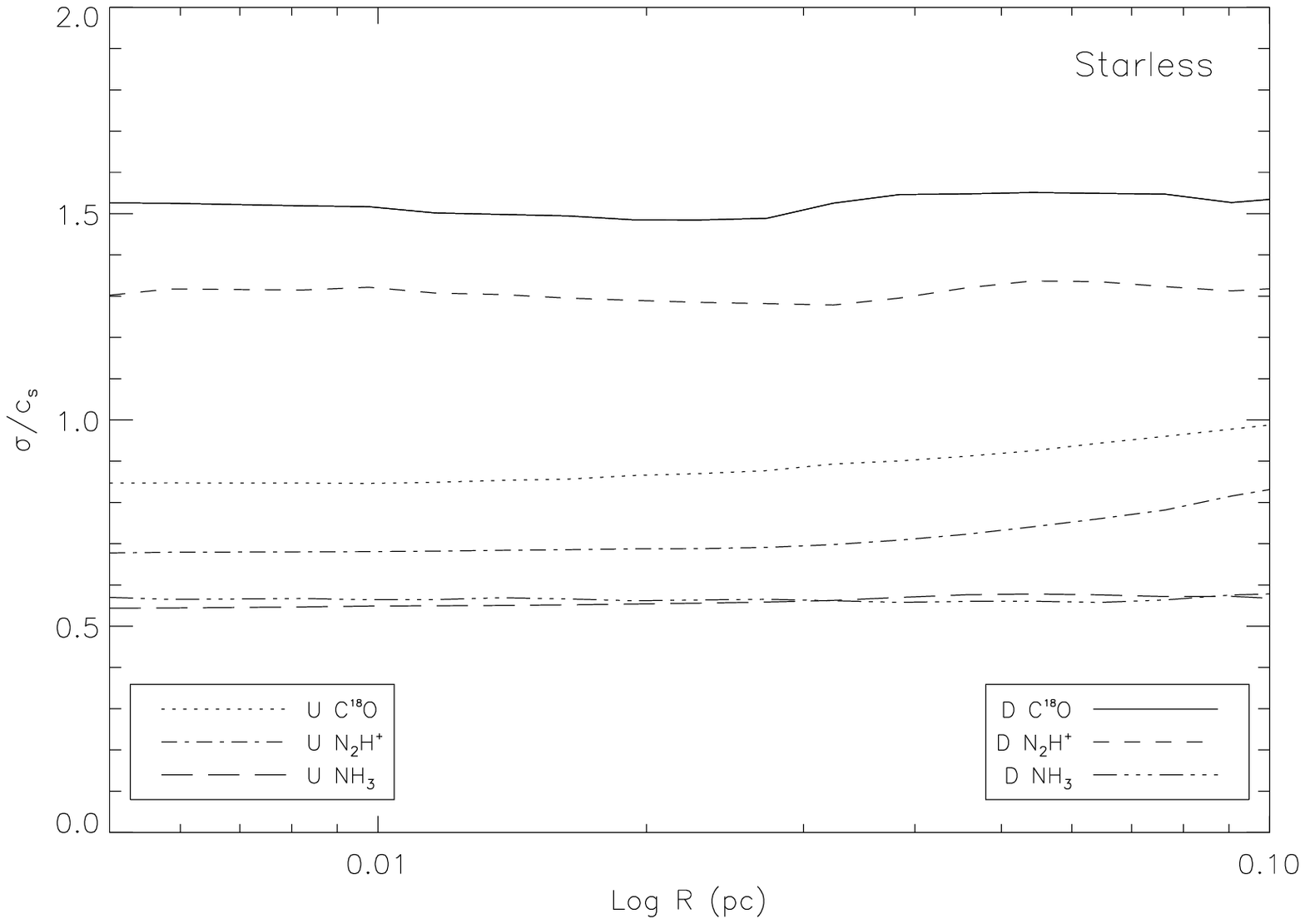} 
\figcaption{The figures show the averaged dispersion of the prestellar cores binned over distance from the central core, where D denotes driven and U denotes undriven turbulence.
\label{ave_vdispr}}
\end{figure*}

\begin{figure*}
\epsscale{0.8}
\plotone{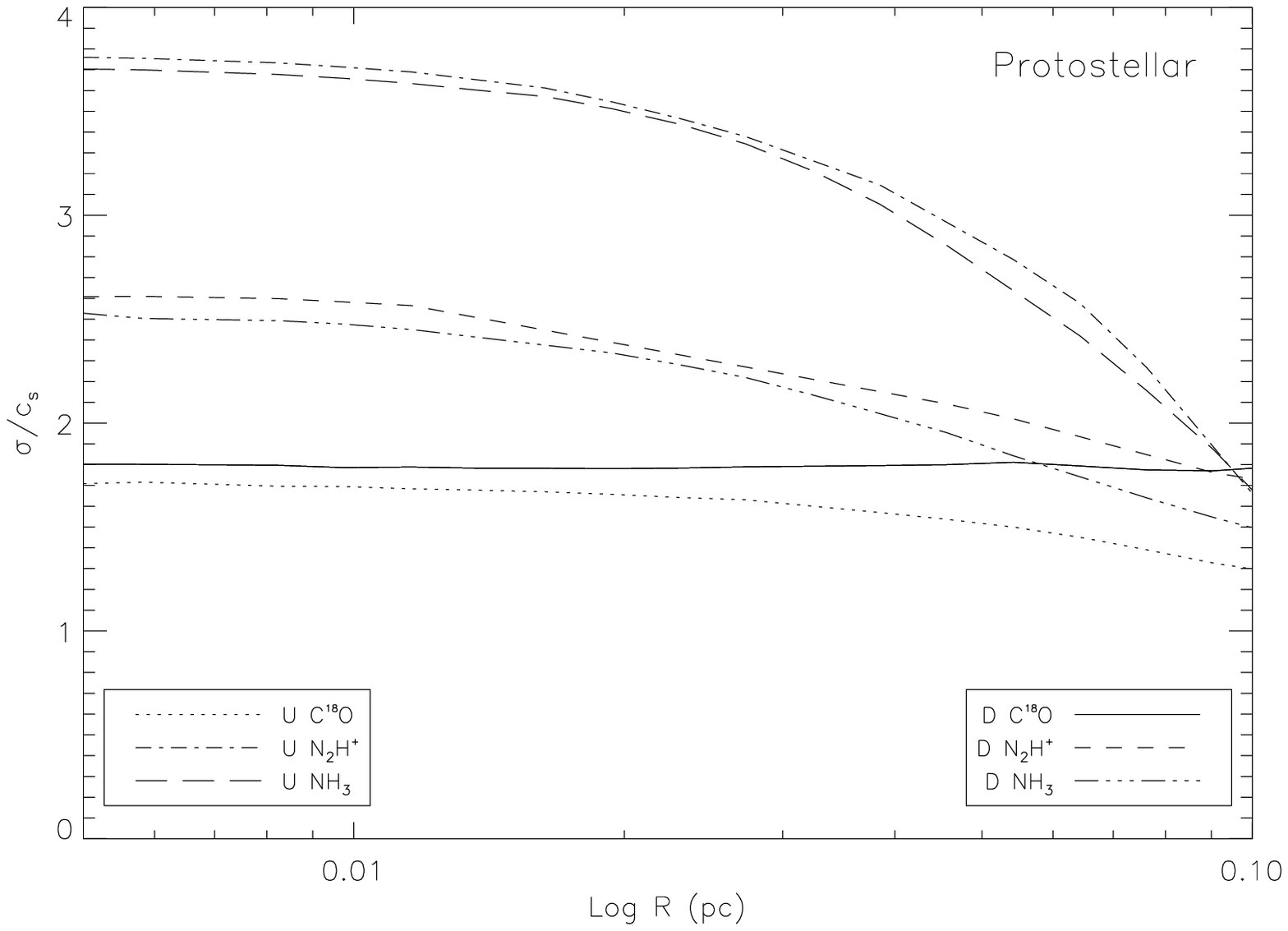}
\figcaption{The figures show the averaged dispersion of only the protostellar cores binned over distance from the central core, where D denotes driven and U denotes undriven turbulence.
\label{ave_vdisprs}}
\end{figure*}

In order to compare the two environments observed with three common tracers, C$^{18}$O, N$_2$H$^+$, and NH$_{3}$, we calculate the velocity dispersion through each pixel along the line of sight.  Figures \ref{dtCO_N2H+} and \ref{vpCO_N2H+} show the velocity dispersion of each pixel in the vicinity of a single prestellar and protostellar core for decaying turbulence, which represent typical examples of each type from our sample, overlaid with contours of integrated intensity. 
The large number of cores in our sample makes comparing the populations by eye on an individual basis difficult. 
In order to consolidate the data sets for each environment,  we bin the pixels by radial distance from the core center. We define 20 logarithmic bins that range from 0.005 to 0.1 pc
in projected distance from the core center and then average together the velocity dispersions of all pixels that fall into a
given bin, including all prestellar or protostellar cores in each case. The result is a single `averaged' core for each tracer and environment. We have plotted this averaged velocity dispersion 
as a function of distance from core center in figures \ref{ave_vdispr} and \ref{ave_vdisprs} for starless and protostellar cores, respectively. There are several interesting points that may be noted from these plots.

First, gas sampled by low density tracers (e.g. C$^{18}$O) around prestellar cores has a higher velocity dispersion than that sampled by higher density tracers. This is reasonable given that the lower-density gas is further from the core center and generally more turbulent. Before collapse ensues, the cores have typically not developed strong high density peaks as is evident in figure \ref{dtCO_N2H+}. This difference between lower and higher density tracers has been frequently exploited observationally to distinguish between the dense core and surrounding envelope (e.g. K07; Walsh et al. 2004).  

Second, figure \ref{ave_vdispr} shows that the starless cores forming in the driven simulation tend to have a higher average velocity dispersion than those in the decaying simulation. This is mainly apparent in the tracer C$^{18}$O, which traces the more turbulent core envelope. 

Most importantly, the average prestellar velocity dispersion for both cases and for all tracers are approximately sonic.  Even the lowest density tracer, C$^{18}$O, remains, on average, below $2c_s$ for the range of column densities in the core neighborhood. 

Finally, we note that there is only a small increase in the dispersion with 
increasing radius. This is consistent with observations by Barranco \& Goodman (1998) and Goodman et al. (1998) who find that the velocity dispersion of the cores on the scale of $\sim$ 0.1 pc is approximately constant, with a small increase near the edge of this region of "coherence." 
The magnitude of the dispersion suggests that the starless cores forming in a turbulent medium are not strongly confined by shocks in the range of densities that are traced by observers.

In contrast, some of these conclusions do not hold for protostellar cores, when strong infall occurs. As shown in figure \ref{ave_vdisprs}, protostellar cores exhibit significantly higher average velocity dispersions than the prestellar counterparts.  The tracers of the protostellar cores behave differently as well. Due to the strong infall, which occurs in the densest gas, the higher density tracers,  N$_2$H$^+$ and NH$_3$, show higher velocity dispersions than the C$^{18}$O, which indicates that the lower density envelope remains transonic. 

There is also clearly a significant difference between the protostellar cores in the two environments.  Those cores in the driven environment have transonic to slightly supersonic velocity dispersions in all tracers that do not vary significantly with
distance from the core center, which is consistent with the coherent core structure observed. This indicates that the cores still have residual turbulent pressure support at a global freefall time and collapse more slowly. However, the protostellar cores in the decaying turbulence environment, lacking this support, have shorter lifetimes and proceed more quickly to collapse and develop much higher, supersonic, central velocity dispersions in N$_2$H$^+$ and NH$_3$ as the cloud gas infalls to the high density regions.  At large radii
however, the velocity dispersion of the protostellar cores in the decaying
enviroment matches the velocity dispersion of cores in the driven environment.
A similar time-dependent trend is obtained in decaying simulations by Ayliffe et al. (2007). 

In summary, prestellar cores forming in driven turbulence have average dispersions of
$\simlt 1.5 c_{\rm s}$ in all tracers, and this dispersion is either flat or
slowly decreasing with increasing radius. In contrast, cores in decaying turbulence show
small ($\sigma_{\rm NT} < 1.0c_{\rm s}$), flat dispersions for prestellar cores, but large
and radially decreasing dispersions for protostellar cores. This is most likely due to infall of unbound gas from large distances at late times, which is a signature of
competitive accretion. We do not observe this in the driven run because
the cloud gas dispersion is too high for Bondi-Hoyle accretion to be
efficient over large distances (Krumholz et al. 2006a).

The dispersions we obtain for the cores and their surrounding envelopes are somewhat dissimilar to those obtained by Klessen et al. (2005) in SPH simulations. As we do, Klessen et al. investigate the velocity dispersions of cores forming in an isothermal, large scale driven turbulent environment. In their study, they derive clump properties when only 5\% of the mass is in cores or at $\sim 0.4t_{\rm ff}$, a much earlier time than we use. However,  even for prestellar cores with driving, they frequently find strong supersonic shocks with $\sigma_{\rm LOS} \sim 3-5c_{\rm s}$ bounding the cores,  which is thus far not supported by observations.  In lieu of a simulated observation, they use a column density cutoff to make the dispersion estimates. We find that we obtain higher velocity dispersions calculating the velocity dispersion directly as Klessen et al. do rather than fitting the line profile in the manner of observers. The reason for the difference is that in some cases the spectra resemble a fairly narrow peak, which is well fit by a Gaussian, surrounded by a much broader base around the 10\% level. The magnitude of this extra spread is reduced substantially at the higher densities as traced by N$_2$H$^+$, and it is likely neglected in the fits performed by observers due to the inherent low-level noise in the actual spectra. Another possibility for the difference is the difficulties of SPH in rendering shocks and instabilities, in particular shear flow instabilities (Agertz et al. 2007) that are likely to be present in any compressible turbulent simulation and may seriously affect accuracy. However, the extent that this may contribute to the high dispersions  found by Klessen et al. is unclear. 

\subsection{Relative Motions}

Observers frequently evaluate an intensity-weighted mean velocity, or first moment, along the line of sight through the core center. While the second moments are indicative of infall motions, the first moments represent the net core advection. The dispersion of the first moments indicates how much the cores move relative to one another.  Observations find that the dispersion of first moments is generally smaller than the velocity dispersion of gas that is not in cores, although how much so varies from region to region.  For example, A07 conclude that the first moment dispersion is sub-virial by a factor of $\sim$ 4 in $\rho$ Ophiuchus.  K07 find that first moment dispersion of starless cores in Perseus is  sub-virial by a factor of $\sim$ 2, which does not rule out virialization. 

 In order to get an unbiased distribution for comparison, it is necessary to subtract out any large gradients in the sample of first moments. Thus, for each region we fit $V = V_0 + \del V \cdot x $ as a function of position, $x$. 
Generally, this turns out to be a fairly small correction, but the net effect is to decrease the dispersion of first moments relative to the gas

\begin{figure*}
\epsscale{0.8}
\plotone{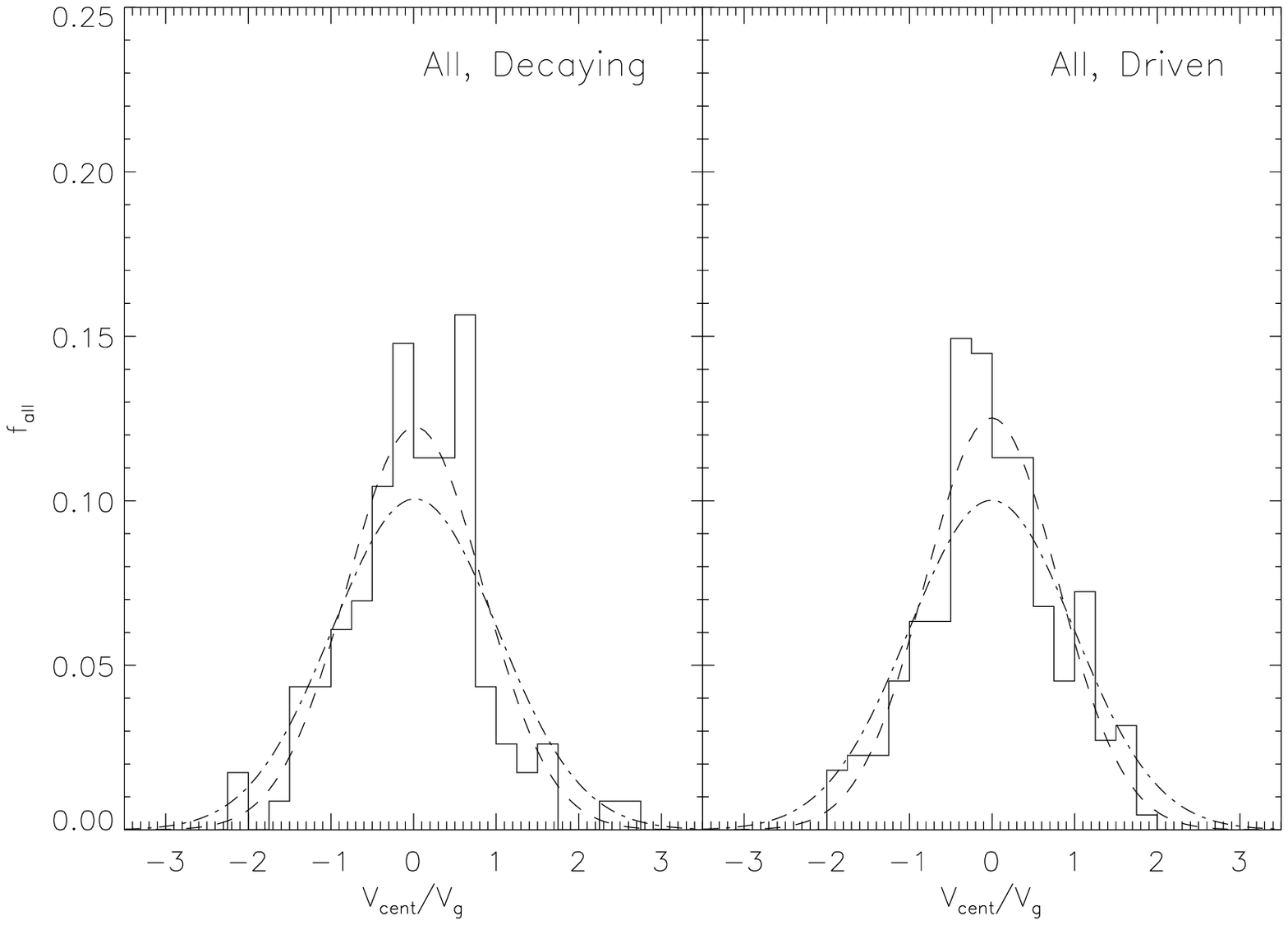}
\figcaption{ Fraction $f$ of all cores binned as a function of first moments, $V_{\rm cent}$, for a simulated observation using N$_2$H$^+$ normalized to the large-scale gas dispersion. $V_{\rm g}$ at $t=t_{\rm ff}$.
The distribution on the left shows the cores in the decaying turbulence environment, while the distribution on the right gives the cores in the driven turbulence environment. The dashed line is a Gaussian with the same dispersion as the data while the dot-dashed line is a Gaussian with the gas velocity dispersion  ($V_{\rm g}=2.2c_{\rm s}$, $V_{\rm g}=4.9c_{\rm s}$, for the decaying and driven simulations, respectively).  \label{allcdisp}}
\end{figure*}

\begin{figure*}
\epsscale{0.8}
\plotone{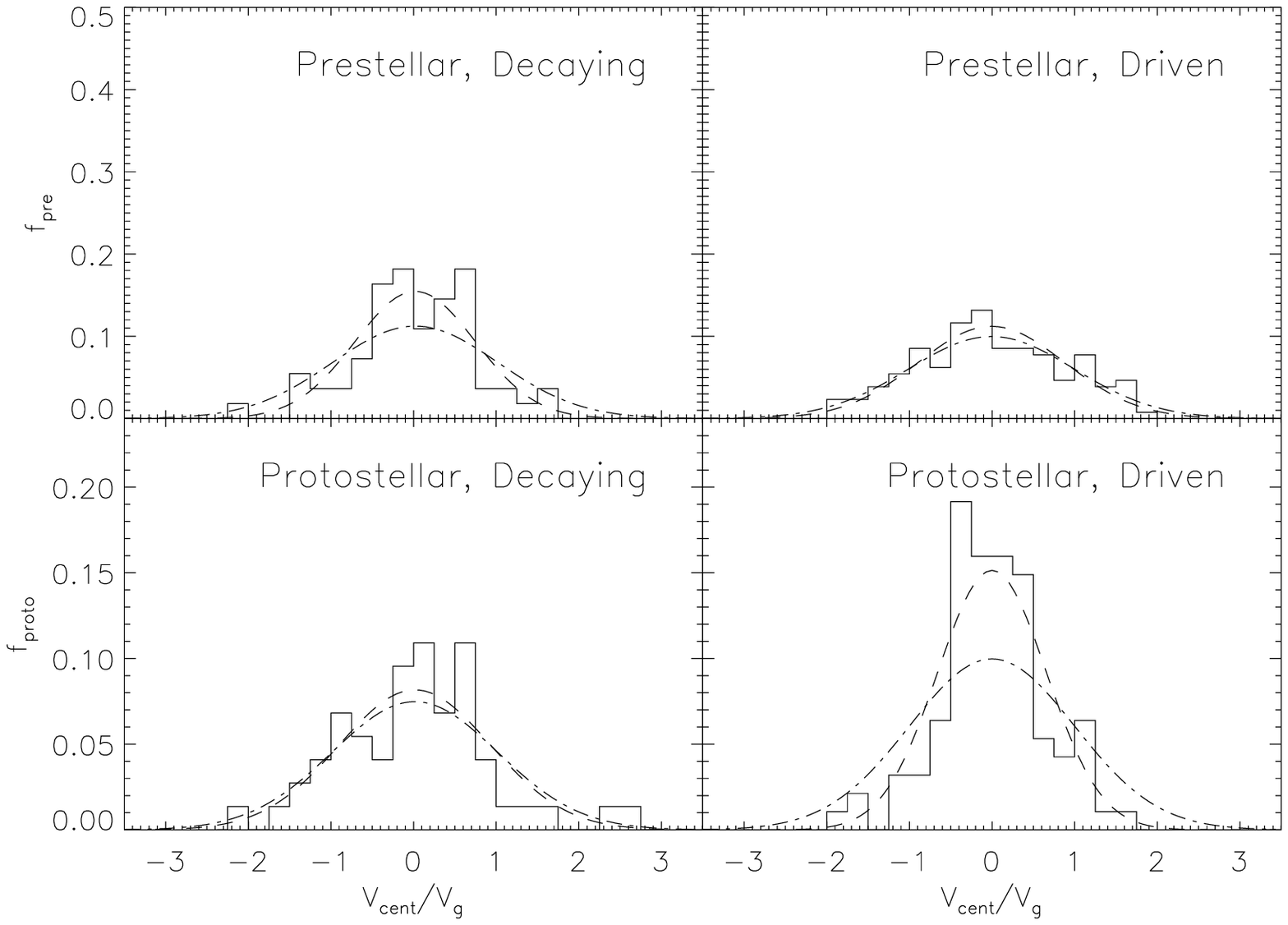}
\figcaption{ Fraction $f$ of prestellar cores (top) and protostellar cores (bottom) binned as a function of first moments, $V_{\rm cent}$, for a simulated observation using N$_2$H$^+$ normalized to the large-scale gas dispersion, $V_{\rm g}$.
The distribution on the left shows those cores in the decaying turbulence enviornment, while the distribution on the right gives the cores in the driven turbulence enviornment. \label{preprocdisp}}
\end{figure*}

 We plot the distribution of first moments for all cores in both environments in figure \ref{allcdisp}, and we plot the distributions for prestellar and protostellar cores separately in figure \ref{preprocdisp}.  In these, we normalize to the ``measured" gas dispersion and correct for the velocity gradient in the box. The dashed line is a Gaussian with the same dispersion as the core distribution. For reference, we also plot a Gaussian with the gas dispersion.  Note that in the driven simulation the dispersion inferred from virial arguments and the time-dependent gas dispersion are the same, because by definition we fix the total kinetic energy to maintain virial balance. However, for the decaying simulation, the time-dependent gas dispersion is lower than would be derived from a virial argument using the total gas mass and cloud size.

Again, we use KS tests to characterize similarity in the populations, which we report in Table \ref{table3}. A KS test indicates that driven and decaying distributions of the net first moments agree with 56\% confidence, while the prestellar and protostellar core first moments agree with 40\% and 13\% confidence. This is significant enough to imply that the early core motions are not widely different in the two environments, with the largest difference occurring between the protostellar first moments. Comparing these distributions with a Gaussian dispersion at the gas dispersion yields good agreement for the distributions of the prestellar driven cores (54\% confidence) and protostellar decaying cores (56\%), but low agreement for the other distributions. In general, low agreement may be because the first moment distributions, although having a similar dispersion to the gas in some cases, are not well represented by a Gaussian distribution.

 In Table \ref{table5}, we list the first-moment dispersions, both corrected and uncorrected for large linear gradients.  We find that the corrected net core dispersion for the driven and decaying cores are both sub-virial relative to the gas dispersion. Previous simulations have shown that the dispersion of first moments becomes sub-virial towards higher gas densities (Padoan et al. 2000), so the result is not unexpected.  One interesting difference between the simulations is that the decaying $\it{protostellar}$ cores are approximately virial, while the $\it{prestellar}$ driven cores are approximately virial.
The former suggests that as the cloud loses turbulent support and tends toward global collapse, that either the core interactions increase or that the cores retain some memory of their natal gas dispersion. The inertia of the cores implies that their velocity dispersions will tend to decay more slowly than that of the gas as a whole. This is a potentially testable signature of the competitive accretion model (Bonnell et al. 2001). In the latter case, the prestellar cores may still be forming out of the shocking gas and hence may still have similar motions. In general, the sub-virial dispersion of the cores may imply that they are not scattering sufficiently frequently to virialize within the formation timescale. Elmegreen (2007) reasons that if cores form at the intersection of two colliding shocks, then their initial dispersion should be on average less than the gas dispersion.  Overall, our results imply that the forming cores are at least somewhat sensitive to the actual dispersion of the natal gas.

\section{ Observational Comparisons}

\subsection{Scaling to Observed Regions}

In this section, we compare our simulated observations with three selections of cores observed in three standard molecular tracers in two different low-mass star-forming regions,
$\rho$~Ophiuchus (primarily L1688) and the Perseus Molecular Cloud.
This comparison cannot be precise for several reasons:
First, the cloud is isolated, whereas our
simulation is a periodic box; second, we are using a single simulation
with given values of the virial parameter and the Mach number to compare
with clouds that have somewhat different values of each of these parameters;
and, finally, our simulation is isothermal, whereas the temperature is observed
to vary in the clouds. Furthermore, the actual cloud is magnetized, whereas
our simulation is purely hydrodynamic. A variety of possible comparison
strategies is possible. We have chosen to use the same mean density in
the box as in the cloud, and to make the simulation temperature agree approximately
with the typical temperature observed in the cloud cores. The size and mass of the 
simulation box then follow from equations (\ref{eq:l}) and (\ref{eq:m}). With this
approach, the Jeans mass will be about the same in the simulation and in the cloud,
but the size and mass of the overall cloud will generally differ between the two.

A07 observed 41 starless cores in $\rho$ Ophiuchus and made maps of 26 of them using the tracer N$_2$H$^+$ ($J=1\rightarrow0$), which are clustered in a region 
of area
1.1 pc$^2$. 
The total gas mass in this region with extinction greater than 15 magnitudes is estimated to be $\sim$ 615 $\msun$ 
(Enoch, private communication; Enoch et al. 2007) with peak column densities of $N_{\rm H_2}$=1-8$ \times 10^{23}$ cm$^{-2}$ (Motte \& Andr\'{e} 1998). 
The star-forming area of $\rho$ Ophiuchus is roughly 
circular
with radius $R\simeq 0.6$ pc;
the mean density and column density are therefore $\nbh\simeq 2\times 10^4$~cm\eee\
and $N_{\rm H}=5\times 10^{22}$~cm\ee.
As discussed above, we adopt this density for our simulation.
To fix the temperature, we first note that
dust temperatures in the pre-stellar cores range from 12-20 K (A07).
On the scale of the entire L1688 cloud, the temperatures as measured by $^{12}$CO 
and $^{13}$CO lines are 29~K and 21~K, respectively (Loren 1989a; in his notation, this
region is R22). We therefore adopt $T=20$~K for the simulation. Equations
(\ref{eq:l}) and (\ref{eq:m}) give $L=0.9$~pc and $M=550\;M_\odot$
for the simulation box, comparable to, although somewhat less than, the observed values.
The total velocity dispersion measured from the $^{13}$CO line
is 1.06~km~s\e\ (Loren 1989b), which lies above the standard
linewidth-size relation (eq. \ref{eq:lws}). The corresponding 1D Mach number is
$\calm_{1D}=3.9$, slightly less than the value 4.9 in the simulation. The
virial parameter of the cloud is 1.25, also slightly less than the simulation value of
1.67.

\begin{deluxetable}{l l l l l}
\tablecaption{KS statistics for the driven and decaying core first moments (centroid velocities) corrected for large velocity gradients and the gas. \label{table3}}
\tablewidth{0pt}
\tablecolumns{5}
\tablehead{
\colhead{} & \colhead{D: All} &
\colhead{D: Starless} & \colhead{D: Proto} &
\colhead{Gas: ${\cal M}_{\rm 1D}$=4.9} }
\startdata
U: All 	 	& 56\% 	& 23\% 	&    44\% 		& 2 \%	 \\  
U: Starless  	& 68\%	&40\%	&   89\%		& 54\% \\ 
U: Proto  & 53\%	& 54\%	&   13\%		& 1 \% 	\\  
Gas: ${\cal M}_{\rm 1D}$=4.9 & 14\%	& 14 \% &  56\%	&  -  
\enddata
\tablecomments{D = driven, U = undriven}
\end{deluxetable}

\begin{deluxetable*}{cccccccccccccc}
\tablewidth{0pt}
\tablecolumns{14}
\tablecaption{Dispersion of first moments (centroid velocities) normalized to the large-scale gas 
dispersion.\label{table5}}
\tablehead{
\colhead{} &
\multicolumn{4}{c}{All} &
\colhead{} &
\multicolumn{3}{c}{Protostellar} &
\colhead{} &
\multicolumn{4}{c}{Prestellar} \\
\colhead{} & \colhead{D} &
\colhead{U} & \colhead{K07} &
\colhead{R07} & \colhead{} &
\colhead{D} & \colhead{U} &
\colhead{K07} & \colhead{} &
\colhead{D} & \colhead{U} &
\colhead{A07} & \colhead{K07}
}
\startdata
$\sigma_{\rm V}/\sigma_{\rm g}\tablenotemark{a} $
& 0.89 & 0.97   & 1.62       & 1.50      & & 0.73   &   1.04  & 1.31    & & 1.00   & 0.90    & 0.75 &  1.81 \\
$\sigma_{\rm V_{\rm cor}}/\sigma_{\rm g}\tablenotemark{b} $
& 0.80 &  0.82  & 1.02     &  0.98       & & 0.66   &  0.92 & 0.98  & & 0.89    & 0.73  & 0.46 & 1.03
\enddata
\tablecomments{D = driven, U = undriven, K07 = Kirk \etal (2007), R07 = Rosolowsky \etal (2007), A07 = Andr\'{e} \etal (2007)
}
\tablenotetext{a}{Uncorrected for linear gradients}
\tablenotetext{b}{Corrected for linear gradients}
\end{deluxetable*}

For the Perseus MC,
K07 report central velocity dispersions and centroid velocities measured from C$^{18}$O and N$_2$H$^+$ pointings for 59 prestellar and 41 protostellar cores.
 R07, also making pointed observations of Perseus,  obtain velocity dispersions and centroid velocities for 199 prestellar and protostellar cores using NH$_3$ (2,2), NH$_3$ (1,1) and C$_2$S (2,1). They adopt a dust temperature of 11 K, which is slightly lower than the assumed temperature of 15 K used by K07.
In comparison to $\rho$ Ophiuchus, the Perseus star-forming region is much larger, 5 pc $\times$ 25 pc, resembles a long chain of clumps with typical column densities of $N_{\rm H_2} \sim 3 \times 10^{22}$ cm$^{-2}$, and contains a total mass of $\sim$ 18,500 $\msun$ (Kirk et al. 2006).  Using the total mass 
and assuming a cylindrical geometry ($L=25$~pc and $R=2.5$~pc) we obtain $\bar n_{\rm H} = 1.1 \times 10^3$ cm$^{-3}$ for Perseus, which we adopt for the simulation.  
We assume that Perseus is approximately in the plane of the sky; if it were randomly oriented then 
the expected  value of
the longest side of the cloud would be 50~pc.
We take a temperature of 10~K for Perseus,
since this is characteristic of the prestellar cores (R07).
Equations (\ref{eq:l}) and (\ref{eq:m}) then imply that the simulation box
has $L=2.8$~pc and $M=825\; M_\odot$, which is a relatively small piece of
the total cloud. Since we are simulating only a small part of the Perseus cloud,
we estimate the velocity dispersion in actual molecular gas from the average
linewidth-size relation (eq. \ref{eq:lws} for $L=5$~pc), which gives $\sigma = 1.1$ km s$^{-1}$
and $\calm_{1D}=5.9$.  
In comparison, our simulation box scaled to the Perseus average number density is 
less turbulent and only half the length of the shorter dimension.
This difference in Mach number and cloud side yields a virial parameter for Perseus of $\alpha \simeq 1$, which is about 60\% of
the value of our simulation box.

\begin{deluxetable}{c c c c c c c }
\tablecaption{Total optical depth $\tau$ through core centers for each normalization and simulated racer. 
\label{optdepth}}
\tablewidth{0pt}
\tablehead{
\colhead{} &
\multicolumn{3}{c}{Perseus} &
\multicolumn{3}{c}{$\rho$ Ophiuchus\tablenotemark{a}} \\
\colhead{$\tau_{\rm tot}$\tablenotemark{b}} &\colhead{median} &
\colhead{min} & \colhead{max} &
\colhead{median} &
\colhead{min} & \colhead{max} 
}
\startdata
C$^{18}$O & 
0.51 & 0.08  & 2.46	 & 0.35 &   0.14 &  1.05 \\
N$_2$H$^+$ & 0.71 &  0.07 &	8.91 & 7.27 &   1.72 &    29.44 \\
NH$_3$ & 8.37 &  0.10 &	63.49 & 46.59 &   10.61 &      228.73 \\
\enddata
\tablenotetext{a}{ Optical depths are reported for the distribution of starless cores only.}
\tablenotetext{b}{$\tau_{\rm tot}$ is the sum of the optical depths through line center for each hyperfine transition. For 
N$_2$H$^+$ and NH$_3$ with 7 and 18 hyperfine transitions, respectively, the optical depth is significantly reduced and generally optically thin for individual transitions. }
\end{deluxetable}
 
\subsection{Optical Depths}
\label{odepths}

In our analysis we make the assumption that the line transitions are optically thin. This approximation is observationally validated for both the N$_2$H$^+$ and NH$_3$ transitions. For example, according to K07 the total optical depth, $\tau_{\rm tot} \sim 0.1-13$, where $\tau_{\rm tot}$ is the sum of the optical depths for each hyperfine transition. Thus, the average optical depth for a given N$_2$H$^+$ hyperfine line is $\bar\tau =\tau_{\rm tot}/7 \sim 0.01- 2$, so that the majority of the lines are at least marginally optically thin. 
In particular, the isolated 101-012 hyperfine component used for velocity fitting has an optical depth of $\tau_{\rm tot}/9$, and is therefore optically thin in all but the very densest cores.
A07 report similar N$_2$H$^+$ total optical depths of $\tau_{\rm tot} \sim 0.1-30$ for $\rho$ Ophiuchus. R07 find $\tau_{\rm tot} \sim 0.4-15$ for NH$_3$. The NH$_3$ (1,1) complex has 18 hyperfine components so that most of the lines are at least marginally optically thin. For comparison, we report the total optical depth in our simulations for all three tracers in Table \ref{optdepth}. 
We derive the optical depth for a given line by solving for the level populations as described in \S 3. Once these are known, the 
opacity in each
cell for photons emitted in the transition from state $i$ to state $j$ is
\begin{equation} \kappa = n X
\frac{f_j B_{ji} \phi(v_{\rm obs};v)} {4 \pi (v_1 - v_0) \nu_{ij}},
\end{equation}
where $n$, $v$, and $X$ are the number density, velocity, and
molecular abundance in the cell, $B_{ji}$ and $\nu_{ij}$ are the
Einstein absorption coefficient and frequency of the transition, and
the observation is made in a channel centered at velocity $v_{\rm
obs}$ that runs from velocity $v_1$ to $v_0$. The optical depth is
given simply by computing this quantity in every cell, multiplying by
the cell length to obtain the optical depth of that cell, and then
summing over all cells along a given line of sight.
As the table shows, for the most part the average hyperfine transition
is optically thin in all tracers. The main exception is cores traced
by NH$_3$ in $\rho$ Ophiuchus, which is marginally optically thick. As
a result, we do not present results for NH$_3$ using the higher
density $\rho$ Ophiuchus scaling; the core velocity dispersion maps in
Figures 5-8 are normalized to Perseus.

\begin{figure*}
\epsscale{0.8}
\plotone{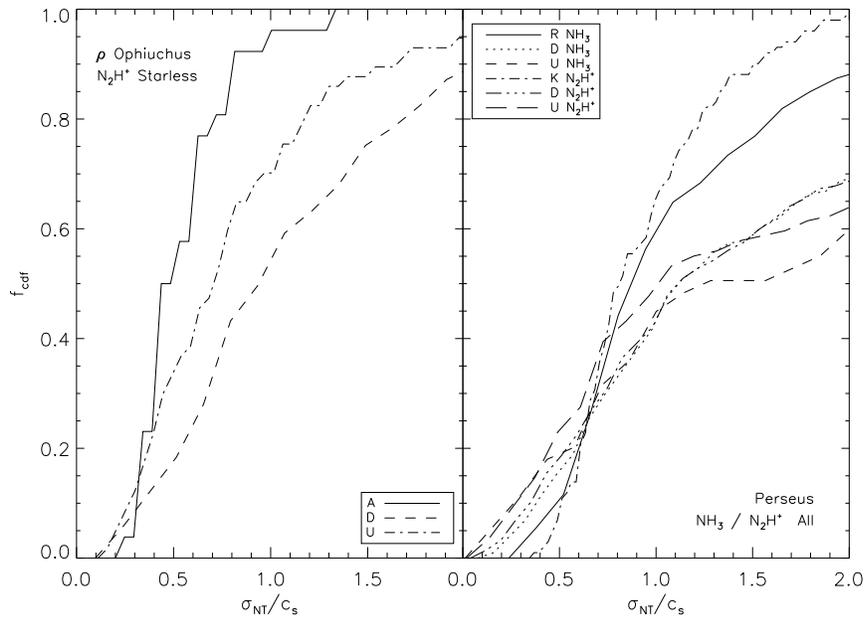}
\figcaption{Cumulative distribution function showing the total fraction $f$ of cores with second moments, $\sigma_{\rm NT}$, less than or equal to the x coordinate value for simulated observations of $\rho$ Ophiuchus and Perseus in N$_2$H$^+$ and NH$_3$.  
The legends indicate by first letter whether the distribution is taken from K07, A07, R07, Undriven simulation, or Driven simulation. The tracer is also indicated when two different tracers are used. 
 \label{dispcdf}}
\end{figure*}

In all other cases even the strongest hyperfine components have
optical depths of order unity, and comparison with more detailed
radiative transfer modeling than we perform indicates this is
unlikely to significantly affect our results. Tafalla et al.\ (2002)
model the emission and transfer of the same N$_2$H$^+$ and NH$_3$
lines that we use in a sample of starless cores in Taurus and
Perseus whose conditions are similar to those produced by our
simulations. They study the effect of the interplay between hyperfine
splitting and radiative trapping by analyzing
the two limiting cases of negligible radiative trapping (which we
assume) and neglect of hyperfine splitting (which maximizes radiative
trapping). They find that the difference in the level populations they
compute under these two assumptions is only a
few tens of percent, a level of error comparable to that introduced by
uncertainties in the collision rate coefficients. We expect the errors
introduced by our optically thin assumption to be comparable.

\subsection{Comparison of Second Moments}

Observationally, the second moments of  cores are predominantly subsonic in MCs, apparently independent of the amount of turbulence. For example, A07, measuring second moments in $\rho$ Ophiuchus, find all values are smaller than $2c_s$ with an average $\sigma_{\rm NT}/c_{\rm s} = 0.5$. Likewise, K07 report similar measurements for  cores observed in Perseus, finding an average of $\sigma_{\rm NT}/c_{\rm s}$= 0.7 with a maximum value of 1.7.
Both our simulations find marginally sub-sonic distributions of second moments with slightly larger means than the observations (see Table \ref{table1}).
In comparison, protostellar cores are observed to have a somewhat broader distribution of second moments. K07 find that the protostellar cores in Perseus have a mean second moment of 1.1$c_s$ and a maximum of 2.3$c_{\rm s}$. The protostellar objects that we observe in our driven simulation tend to have transonic second moments while in the decaying simulation they are supersonic. 

\begin{deluxetable}{cccc}
\tablecaption{KS statistics for the driven and decaying core second moments (velocity dispersions) compared to the observational collections of cores using the appropriate cloud normalization and simulated tracer. 
\label{table2}}
\tablewidth{0pt}
\tablecolumns{3}
\tablehead{
\colhead{Sample} &\colhead{Cloud} &
\colhead{D} & \colhead{U}
}
\startdata
Starless & $\rho$ Ophiuchus (A07)         & 8x10$^{-4}$\%  &  2\%    \\  
& Perseus (K07)     &  2\%& 2x10$^{-2}$\%   \\
Protostellar & Perseus (K07)      & 2x10$^{-4}$\%& ... \\
All & Perseus (K07)      & 1x10$^{-3}$\% & 8x10$^{-4}$  \\  
& Perseus (R07)      & 1\% & ...  
\enddata
\end{deluxetable}

We use a KS test to compare the distribution of second moments for each of the simulation core populations with the observed core populations. We give the results in Table \ref{table2}. Note that the A07 sample is comprised of only prestellar cores, while R07 observe both starless and protostellar cores but do not distinguish between them. Figure \ref{dispcdf} shows the cumulative distribution functions of the core populations for some of the simulations and observations. Although the medians of some of the second-moment distributions are fairly similar, KS tests of the core populations show significant disagreement in some cases. Overall, the distribution of second moments for the driven run is closer to observations of Perseus, while the decaying run is a better match for the $\rho$ Opiuchus prestellar second moments. 

The physical origin of the poor agreement between the simulations and observations appears to be that the simulated protostellar second-moment distributions in either case do not have sufficiently narrow peaks. The protostellar cores in the simulations are at the centers of regions of supersonic infall, which contradicts the observations that show at most transonic contraction. 
Although the decaying simulation has a larger population of high dispersion protostellar cores, both simulations show almost equally bad agreement with the observations.
Tilley \& Pudritz (2004), performing smaller decaying turbulent cloud simulations at lower resolution with self-gravity, analyze the linewidths of their cores using a similar simple chemical mode. They also find a number of cores with greater than sonic central linewidths.
There are two possibilities for the discrepancy between the observed protostellar cores in our simulation and those observed in Perseus. In reality, forming stars are accompanied by strong outflows that may eject a large amount of mass from the core, leading to efficiency factors between $\epsilon_{core}=$0.25-0.75 (Maztner \& McKee 2000). Such outflows limit the mass of the forming protostar by this amount. Since we do not include outflows we naturally expect our sink particles to overestimate the forming protostar mass by this factor and hence the maximum infall velocity, characterized by the second moment through core center. If we adopt a sink particle mass correction of 3 (Alves et al. 2007),  then the infall velocity will decrease by a factor of $\sqrt{3}$.
This correction substantially reduces the number of protostellar cores with supersonic second moments from 53\% and 70\% to 23\% and 39\% for cores in the driven and decaying simulations, respectively. This correction brings the driven core sample closer in agreement with those measured by R07 and K07. A second possibility for the higher second moments is the lack of magnetic fields in our simulations. Magnetic pressure support could also retard collapse and decrease the magnitude of the infall velocities. However, the importance of magnetic effects is difficult to assess without further simulations.

\begin{figure*}
\epsscale{0.8}
\plotone{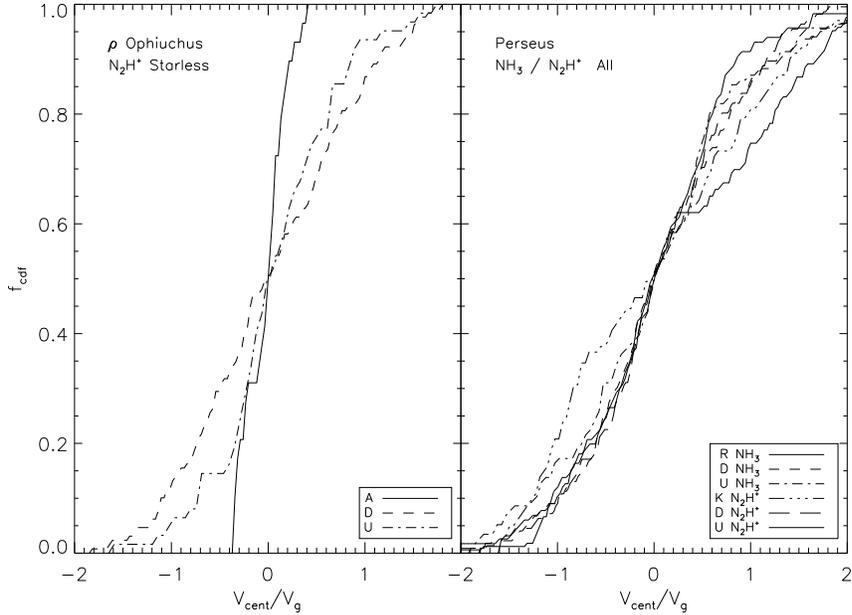}
\figcaption{Cumulative distribution function showing the total fraction $f$ of cores with first moments, $V_{\rm cent}$, less than or equal to the x coordinate value for simulated observations of $\rho$ Ophiuchus and Perseus in N$_2$H$^+$ and NH$_3$.  Each line is normalized to the appropriate large-scale gas dispersion, $V_{\rm g}$, either as measured (simulations) or as derived from the linewidth-size relation in equation (6).
The legend format is similar to figure \ref{dispcdf}.
\label{centcdf}}
\end{figure*}

\subsection{Comparison of First Moments}

 In contrast, we find better agreement between simulations and observations for bulk core motions. When comparing the distributions of first moments, we first subtract out any large gradients in the sample as discussed in \S 4.3. This is particularly important when comparing to a large elongated cloud such as Perseus.  
 We then shift the distributions so that median centroid velocity falls at 0 and normalize the distribution to the bulk gas dispersion. 
 For Perseus, we infer the bulk gas velocity dispersion for our simulation $\sigma = 1.1$ km s$^{-1}$ by assuming the cloud falls on the linewidth-size relation and satisfies equation (6)
with $L$ equal to the transverse size of the cloud.  
 For $\rho$ Ophiuchus, we adopt the  $^{13}$CO line velocity dispersion of $\sigma = 1.06$~km~s\e\ (Loren 1989b).

 \begin{deluxetable}{c c c c  }
\tablecaption{KS statistics for the driven and decaying core first moments (centroid velocities) compared to the observational collections of cores using the appropriate cloud normalization and simulated tracer. 
\label{table4}}
\tablewidth{0pt}
\tablehead{
\colhead{Sample} &\colhead{Cloud} &
\colhead{D} & \colhead{U}
}
\startdata
Starless & $\rho$ Ophiuchus (A07)      & 0.5 \% &  6\% \\  
& Perseus (K07)     & 48\%& 12\%\\  
Protostellar & Perseus (K07)     & 6\%& 85\%\\  
All & Perseus (K07)     & 0.8\%& 7\%\\  
& Perseus (R07)     & 7\%&  3\%\\  
\enddata
\end{deluxetable}

In Table \ref{table4}, we report the KS agreement for the first moments of the observations and simulations. Since the simulations themselves are statistically similar to one another, both of the first moment distributions generally either agree or disagree with the observed population. Except in the case of the N$_2$H$^+$ driven data for $\rho$ Ophiuchus and the NH$_3$ decaying data, the velocity-corrected data are fairly statistically similar to the observations. This suggests that the first-moment distributions do not strongly depend upon the details of the turbulence. In figure \ref{centcdf}, we have plotted the cumulative distribution function of some of the first-moment distributions for comparison. The net core distributions show substantial overlap for both simulations and observational regions. The main source of disagreement with observations is the generally larger dispersions of the first moments in the simulations. In particular, the dispersion of the prestellar core first moments is a factor of $\sim 2$  larger than the that found by A07 in $\rho$ Ophiuchus. However, because in both simulations the core-to-core velocity dispersion is smaller than the virial velocity of the cloud on large scales, we conclude that a sub-virial dispersion of first moments is not necessarily an indicator of global collapse.  

 In some cases, the direct dispersion of the gas may be poorly observationally constrained and so a virial argument is used to infer the gas dispersion. We find that normalizing the distributions to the virial gas dispersion rather than the measured gas dispersion produces a significantly different result for the decaying simulation. Since the cloud gas is becoming more quiescent with time, the actual gas dispersion is sub-virial at late times. Thus, relative to the virial gas dispersion the decaying dispersion of first moments appears twice as sub-virial. 
 
\section{ Discussion and Conclusions}

We use isothermal AMR simulations to investigate the kinematics of cores in environments with and without driven turbulence. We simulate observations of these cores in the tracers C$^{18}$O, N$_2$H$^+$, and NH$_{3}$ for the star-forming regions $\rho$ Ophiuchus, 125 pc distant, 
and Perseus, 260 pc distant, with beam sizes of $26"$ and $31"$, respectively. From the differences between cores in the two environments and in conjunction with observational results, we are able to draw a number of important conclusions, some of which are relevant for observationally distinguishing between driven and decaying turbulence in star-forming clouds.

We find that in both simulated environments the prestellar second-moment distribution is fairly narrow and peaked about the sound speed. Significant broadness of the protostellar second moment distributions is due to strong infall, such that many cores have central dispersions exceeding $2c_{\rm s}$.  Despite these commonalities, a KS test indicates that the driven and decaying prestellar and driven and decaying protostellar populations are dissimilar to one another. 
In contrast to the second moments, a KS test indicates that the first-moment distributions in the two environments have some overlap: 13\% confidence for protostellar cores and 44\% confidence for prestellar cores. This similarity is an indication that the bulk core advection is decoupled from the gas motions inside the core. The similarity of the KS tests suggests that core first moments are not a good method for distinguishing the two environments.

Examining the gas dispersion in the core neighborhoods reveals interesting differences in the two simulations. We find that by the end of a global freefall time the averaged velocity dispersion increases strongly towards the core center for decaying protostellar cores. However, for decaying prestellar cores and all driven cores this trend is fairly flat or slightly increasing. Thus for both phases the driven cores are coherent, similar to observed cores (Kirk et al. 2007; Barranco \& Goodman 1998; Goodman et al. 1998), while the supersonic velocities observed in decaying protostellar cores are inconsistent with observations.  Thus, investigating the radial dispersion of protostellar cores may make it possible to discriminate between clouds with and without active turbulent energy injection.

We find that the majority of the combined prestellar and protostellar distribution of second moments through the core centers for both environments are below 2c$_{\rm s}$, which agrees with the results of A07 and K07. 
However, neither prestellar core distribution shows a significant confidence level of agreement with the observations.  

As shown in Table 5, 
we obtain sub-virial dispersions of the first moments for both total core populations like A07,  
however our core-to-core dispersions are approximately a factor of 2 closer to virial. 
Although both runs produce sub-virial core-to-core dispersions, we have not shown that either driven turbulence or the small virial parameter of decaying turbulence can produce $\alpha_{\rm vir}$ as small as that found by A07. 

One interesting finding is that the protostellar cores in the decaying run have a core-to-core dispersion that is higher than the gas dispersion measured after a free-fall time. This is a result of the significantly larger dispersion of the protostellar cores compared to the prestellar cores, which may be a result of either increased scattering or of memory of the natal higher dispersion gas. This is in contrast to the driven prestellar cores, which have nearly the same dispersion as the gas, and the driven protostellar cores, which have a sub-virial dispersion. Thus, comparing the starless and protostellar core first-moment dispersion to the net gas dispersion is potentially a means for distinguishing the two environments.
  
An effect that we cannot rule out is the importance of magnetic fields, which we do not treat in our simulations. In addition to seeding the initial clump mass spectrum, the turbulence in our simulations provides support against the cloud's self-gravity, a role that could be filled by either sustained turbulence or magnetic fields or both. 
The very small number of cores observed with supersonic second moments indicates that these cores are collapsing very slowly, a condition that we find is promoted by turbulent support but not throughout the entire core collapse process. 
At present, little computational work has been done to study line profiles for turbulent cores with magnetic fields. Tilley \& Pudritz (2007) present central line profiles for a few cores formed in self-gravitating magneto-hydrodynamic cloud simulations but do not have many statistics. Our simulations also neglect protostellar outflows, which may have an effect on the total core mass and hence the velocity dispersion of the infalling gas in the core center.
  
Another possible source of the quantitative disagreement between observations and our simulations is geometry.  Periodic boundary conditions may do a poor job representing whole, pressure confined molecular clouds. Certainly, the star-forming region of Perseus is more filamentary than round. Further, the cloud Mach numbers for both regions are somewhat uncertain, and it may be necessary to match the Mach number of the simulation to the cloud more exactly to get better quantitative agreement.
 
Overall, we find that the driven simulation agrees better with the cores in Perseus, while the decaying simulation agrees slightly better with the pre-stellar cores in $\rho$ Ophiuchus (our data do
not include protostellar cores there).
Our results indicate that the decaying simulation produces a population of protostellar cores with supersonic velocity dispersions that is largely inconsistent with
the observations of protostellar cores in Perseus.
To reach a firmer conclusion on the validity of driven or decaying turbulence will require more complete data on a larger sample of clouds as well as simulations that allow for magnetic fields, outflows, and 
thermal feedback from the protostars.
 
 \acknowledgments{
 We thank P. Andr\'{e}, D. Johnstone, E. Rosolowsky, M. Enoch and H. Kirk for helpful discussions of their observations.  
Support for this work was provided under the auspices of the US Department
of Energy by Lawrence Livermore National Laboratory under contacts
B-542762 (S.S.R.O.) and DE-AC52-07NA27344 (R.I.K.); NASA through Hubble Fellowship grant HSF-HF-01186 awarded by the Space Telescope Science Institute, which is operated by the Association of Universities for Research in Astronomy, Inc., for NASA, under contract NAS 05-26555 (M. R. K.); NASA ATP grants NAG 05-12042 and NNG 06-GH96G (R. I. K. and C. F. M.), and the National Science Foundation under Grants No. AST-0606831 and PHY05-51164 (C. F. M. and S. S. R. O). Computational resources were provided by the NSF San Diego Supercomputing Center through NPACI program grant UCB267; and the National Energy Research Scientific Computer Center, which is supported by the Office of Science of the U.S. Department of Energy under contract number DE-AC03-76SF00098, though ERCAP grant 80325.}

\begin{appendix}

\section{Statistical Equilibrium for Molecules with Hyperfine
Structure}
\label{hfappendix}

As discussed in Tafalla et al.\ (2002) and Keto et al.\ (2004),
hyperfine splitting in a molecule introduces two complications on top
of the standard calculation of statistical equilibrium. First,
hyperfine splitting of a transition reduces its optical depth by
breaking the line into multiple components. The frequency separation
between the components means that photons generated by a transition
from level $i\alpha$ to level $j\beta$, where the Roman index
refers to the parent level and the Greek to its hyperfine sublevel, 
have a reduced probability of being resonantly absorbed by molecules
in state $j$ that are not in hyperfine sublevel $\beta$. Under our
assumption that all components are optically thin, however, this
effect is not significant. We discuss the extent to which this
approximation holds, and how our results might be modified in cases
where it fails, in \S~\ref{odepths}.

A second, practical complication is that collision rate coefficients
between different hyperfine sublevels are generally unknown. Only the
total rate coefficients summing over all hyperfine states are
known. This makes it impossible to perform a true statistical
equilibrium calculation without introducing additional assumptions,
the most common of which is that the individual hyperfine sublevels
are simply populated in proportion to their statistical
weights. Observations along some sightlines show that this
approximation generally holds for NH$_3$ and that deviations from it
for N$_2$H$^+$ are only of order 10\% (Tafalla et al. 2002; Keto et
al. 2004).

Under the assumption of an optically thin gas, the equation of
statistical equilibrium for a molecular species with hyperfine
structure is
\begin{eqnarray}
\lefteqn{
\sum_j \sum_{\beta} (n_{\rm H_{\rm 2}} q_{j\beta i\alpha} + A_{j\beta
i\alpha} + B_{j\beta i\alpha} I_{\rm CMB}) f_{j\beta}
}
\nonumber
\\
\label{hf_equilibrium}
& = &
\left[\sum_k \sum_{\beta} (n_{\rm H_{\rm 2}} q_{i\alpha k\beta} +
A_{i\alpha k\beta} + B_{i\alpha k\beta} I_{\rm CMB})\right] f_{i\alpha} \\
\label{hf_normalization}
\sum_i \sum_{\alpha} f_{i\alpha} & = & 1,
\end{eqnarray}
where a set of four subscripts $i\alpha j\beta$ indicates a transition
from state $i$, hyperfine sublevel $\alpha$ to state $j$, hyperfine
sublevel $\beta$. The assumption that the hyperfine sublevels are
populated in proportion to their statistical weight then enables us to
write
\begin{equation}
f_{i\alpha} = \frac{g_{i\alpha}}{g_i} f_i,
\end{equation}
where $g_{i\alpha}$ is the statistical weight of sublevel $i\alpha$,
$g_i=\sum_{\alpha} g_{i\alpha}$ is the total statistical weight of all
hyperfine sublevels of level $i$, and $f_i=\sum_{\alpha} f_{i\alpha}$
is the fraction of molecules in any of the hyperfine sublevels of
level $i$. If we make this substitution in equations
(\ref{hf_equilibrium}) and (\ref{hf_normalization}), then they become
\begin{eqnarray}
\label{hf_equilibrium1}
\lefteqn{
\sum_j \sum_{\beta} \left[(n_{\rm H_{\rm 2}} q_{j\beta i\alpha} + A_{j\beta
i\alpha} + B_{j\beta i\alpha} I_{\rm CMB}) \frac{g_{j\beta}}{g_j}
\right] f_j
}
\nonumber
\\
& = & 
\left[\sum_k \sum_{\beta} (n_{\rm H_{\rm 2}} q_{i\alpha k\beta} +
A_{i\alpha k\beta} + B_{i\alpha k\beta} I_{\rm CMB})
\frac{g_{i\alpha}}{g_i} \right] f_i \\
\sum_i f_i & = & 1.
\end{eqnarray}

If the hyperfine sublevels of state $i$ are populated in proportion to
their statistical weight, then the total transition rate from all
hyperfine sublevels of state $i$ to any of the sublevels of state $j$
are given by
\begin{eqnarray}
\label{hfsums1}
q_{ij} & \equiv & \sum_{\alpha} \sum_{\beta} \frac{g_{i\alpha}}{g_i}
q_{i\alpha j\beta} \\
A_{ij} & \equiv & \sum_{\alpha} \sum_{\beta} \frac{g_{i\alpha}}{g_i}
A_{i\alpha j\beta} \\
B_{ij} & \equiv & \sum_{\alpha} \sum_{\beta} \frac{g_{i\alpha}}{g_i}
B_{i\alpha j\beta}.
\label{hfsums3}
\end{eqnarray}
Now note that (\ref{hf_equilibrium1}) represents one independent
equation for each state $i$ and each of its hyperfine sublevels
$\alpha$. If we fix $i$ and add the equations for each hyperfine
sublevel $\alpha$, then equation (\ref{hf_equilibrium1}) simply
reduces to
\begin{equation}
\sum_j (n_{\rm H_{\rm 2}} q_{ji} + A_{ji} + B_{ji} I_{\rm CMB}) f_j =
\left[\sum_k (n_{\rm H_{\rm 2}} q_{ik} + A_{ik} + B_{ik} I_{\rm
CMB})\right] f_i,
\end{equation}
the same as the equation for an optically thin molecule without
hyperfine splitting, provided that the rate coefficients are
understood to be summed over all hyperfine sublevels.

\end{appendix}


\bigskip
\begin{references}{}


\reference{} Agertz,  O., Moore, B., Stadel, J., Potter, D., Miniati, F., Read, J., Mayer, L., Gawryszczak, A., Kravtsov, A., \& Nordlund, . MNRAS, 830, 963

\reference{} Alves, J., Lombardi, M., \& Lada, C.J. 2007, A\&A, 642, 17

\reference{} Andr\'{e}, P., Bolloche, A., Motte, F., \& Peretto, N. 2007, A\&A, 472, 519

\reference{} Ayliffe, B. A., Langdon, J. C., Cohl, H. S., \& Bate, M. R. 2007, MNRAS, 374, 1198

\reference{} Ballesteros-Paredes, J. Klessen R. S., \& V\'{a}zquez-Semadeni. 2003, ApJ, 592, 188

\reference{} Barranco, J.A. \& Goodman, A. A. 1998, ApJ 504, 207


\reference{} Bonnell I.A., Clarke, C.J., Bate, M.R., \& Pringle, J.E. 2001, MNRAS, 368, 573  

\reference{} Dib, S., Kim, J., V\'{a}zquez-Semadeni, E., Burkert, A., \& Ahadmehri, M. 2007, ApJ, 61, 262

\reference{} Elmegreen, B. G. 2007, ApJ, 668, 1064 

\reference{} Elmegreen, B. G. 2000, MNRAS, 530, 277

\reference{} Enoch, M. L., Glenn, J., Evans II, N. J., Sargent, A. I., Young, K. E., \& Huard, T. L 
ApJ. 2007, 666, 982

\reference{} Enoch, M. L., Young, K. E., Glenn, J., Evans II, N. J., Golwala, S., Sargent, A. I., Harvey, P., Aguirre, J., Goldin, A., Haig, D., Huard, T. L., Lange, A., Laurent, G., Maloney, P.,  Mauskopf, P.,  Rossinot, P.,  \& Sayers. J. 2006, 638, 293

\reference{} Goodman, A. A., Barranco, J. A., Wilner, D. J., \& Heyer, M. H. 1998, ApJ, 504, 223
\reference{} Hartmann, L. 2001, AJ 121, 1030

\reference{} Heyer, M. H. \& Brunt, C. M. 2004, ApJ, 615, 45

\reference{} Keto, E. \& Field, G, 2005, ApJ, 635, 1151

\reference{} Keto, E., Rybicki, G.~B., Bergin, E.~A., \& Plume, R. 2004, ApJ, 613, 355

\reference{} Kirk, H., Johnston, D., \& Di Francesco, J. 2006, 646, 1009

\reference{} Kirk, H., Johnston, D., \& Tafalla, M. 2007, 671, 1820

\reference{} Klein, R. I. 1999, JCAM, 109, 123


\reference{} Klessen, R. S., Ballesteros-Paredes, J,  Vazquez-Semadeni, E, \& Durian-Rojas, 2005, ApJ, 620, 786


\reference{krumholz2007a} Krumholz, M. R., Klein, R.I. \& McKee C. M. 2007a, ApJ, 665, 478

\reference{krumholz2007b} Krumholz, M. R., Klein, R.I. \& McKee C. M. 2007b, ApJ, 656, 959

\reference{} Krumholz, M. R., Klein, R.I. \& McKee C. M. 2004, ApJ, 611, 399

\reference{} Krumholz, M. R., Mazner, C. D.,\& McKee, C. F. 2006b, ApJ, 653, 361

\reference{} Krumholz, M. R., McKee, C. F. \& Klein, R. I. 2006a, ApJ, 638, 369

\reference{} Krumholz, M. R., McKee, C. F. \& Klein, R. I. 2005, Nature, 438, 322

\reference{} Krumholz, M. R., \& Tan, J. C.  2007, ApJ,  656, 959

\reference{} Loren, R.B. 1989a ApJ, 338, 902

\reference{} Loren, R.B. 1989b ApJ, 338, 925

\reference{} Nakamura, F \& Li, Z-Y.  2007 ApJ, 662, 395

\reference{} Matzner, C. D. \& McKee, C.F. 2000, ApJ, 545, 364

\reference{} McKee, C. F. 1999, NATO ASIC Proc. 540: The Origin of Stars and Planetary Systems, 29

\reference{} McKee, C. F. \& Ostriker, E. C. 2007 ARA\&A, 45, 565

\reference{} McKee, C. F. \& Tan, J. C. 2002, Nature, 416, 59

\reference{} Motte, F., Andr\'{e}, P., \& Neri, R. 1998, A\&A, 336, 150

\reference{} Muench, A. A. Lada, C. J., Rathborne, J. M,  Alves, J. F., \& Lombardi, M. 2007, accepted to ApJ

\reference{} Offner, S. S. R., Klein, R. I., \& McKee, C. F. 2008, AJ, in press

\reference{} Padoan, P., Juvela, M., Bally, J. \
\& Nordlund \AA. 2000, ApJ, 529, 259

\reference{} Padoan, P., Juevela, M., Goodman, A. A.\& Nordlund,  \AA. 2001, ApJ, 553, 227

\reference{} Padoan, P., \& Nordlund, \AA. 2002, ApJ, 576, 870
	
\reference{} Rosolowsky, E. W., Pineda, J. E., Foster, J.B., Borkin, M. A., Kauffmann, J., Caselli, P., Myers, P.C., \&  Goodman, A. A. 2007, ApJS, 175, 509
	
\reference{schoier2005} {Sch{\"o}ier}, F.~L., {van der Tak}, F.~F.~S.,{van Dishoeck}, E.~F. \& 
	{Black}, J.~H. 2005, AAP, 432, 369

\reference{} Shu, F. H., Adams, F. C., \& Lizano, S. 1987, ARA\&A, 25, 23

\reference{} Solomon, P. M., Rivolo, A. R., Barrett, J., \& Yahil, A. 1987, ApJ, 319, 730


\reference{} Tafalla, M., Myers, P. C., Caselli,  P., Walmsley, C. M.  \&  Comito. C. 2002, ApJ, 569, 815

\reference{} Tafalla, M., Myers, P. C., Caselli,  P., \& Walmsley, C. M. 2004a, A\&A, 416, 191

\reference{} Tafalla, M., Myers, P. C., Caselli, P., \& Walmsley, C. M. 2004b, Ap\&SS, 292, 347

\reference{} Tilley, D. A. \& Pudritz, R. E. 2004, ApJ, 593, 426.
	
\reference{} Tilley, D. A. \& Pudritz, R. E. 2007, MNRAS, 382, 73


\reference{} Truelove, J. K., Klein, R. I., McKee, C. F., Holliman, J. H. II, Howell, L. H., \& Greenough, J. A. 1997, ApJ, 289, L179

\reference{} Truelove, J. K., Klein, R. I., McKee, C. F., Holliman, J. H. II, Howell, L. H., Greenough, J. A., \& Woods, D. T. 1998, ApJ, 495, 821

\reference{} Walmsley, C. M., Flower, D. R., \& Pineau des For{\^e}ts, G. 2003, A\&A, 418, 1035

\reference{} Walsh, A. J. \& Myers, P. C. 2007, ApJ 614, 194

\reference{} Walsh, A. J., Myers, P.C. \& Burton, M. G. 2004 ApJ, 614, 194

\reference{} Young, K. E., Enoch, M. L., Evans II, N. J., Glenn, J., Sargent, A., Huard, T., Aguirre, J., Golwala, S., Haig, D.,  Harvey, P., Laurent, G., Mauskopf, P.,  \& Sayers J. 2006, ApJ, 644, 326 

\end{references}
\end {document}